\documentclass[12pt]{article}
\usepackage{authblk}
\usepackage{here}
\usepackage{amsmath,amssymb,graphics}
\usepackage{bm}
\usepackage{graphicx}           
\usepackage{ascmac}
\usepackage{color}
\usepackage{braket}
\usepackage{here}
\usepackage{cite}
\usepackage[top=25truemm,bottom=30truemm,left=20truemm,right=20truemm]{geometry}
\usepackage{ytableau}
\usepackage{multirow}
\usepackage{afterpage}
\usepackage{cases}
\usepackage{fancybox}
\usepackage[driverfallback=dvipdfmx]{hyperref}
\usepackage{simplewick}
\usepackage{caption}
\usepackage{subcaption}
\hypersetup{colorlinks=true, citecolor=blue, linkcolor=blue, linkbordercolor=0 0 1}
\makeatletter

\@addtoreset{equation}{section}
\makeatother
\newlength{\myh}
\baselineskip=\normalbaselineskip
\newcommand{\beq}{\begin{eqnarray}}
\newcommand{\eeq}{\end{eqnarray}}
\newcommand{\nn}{\nonumber \\}

\def\matt[#1,#2,#3,#4]{\left(%
\begin{array}{cc} #1 & #2 \\ #3 & #4 \end{array} \right)} 

\begin{document}

\begin{flushright}
\hfill{YITP-22-08}~~~~\\
\hfill{KUNS-2931}~~~~
\end{flushright}
\begin{center}
\vspace{2ex}
{\Large \textbf{
Nucleon D-term in holographic QCD
}}

\vspace*{5mm}
\text{Mitsutoshi Fujita}$^a$\footnote{e-mail:
 \texttt{fujita@mail.sysu.edu.cn}},
 \text{Yoshitaka Hatta}$^{b,c}$\footnote{e-mail:
 \texttt{yhatta@bnl.gov}}, 
\text{Shigeki Sugimoto}$^{d,e,f}$\footnote{e-mail:
 \texttt{sugimoto@gauge.scphys.kyoto-u.ac.jp}}
~and~
\text{Takahiro Ueda}$^{g}$\footnote{e-mail:
 \texttt{tueda@st.seikei.ac.jp}}

\vspace*{4mm}

\hspace{-0.5cm}
\textit{{$^a$ School of Physics and Astronomy, Sun Yat-Sen University,  Guangzhou 519082, China
}}\\
\textit{{$^b$ Physics Department, Brookhaven National Laboratory, Upton, NY 11973, USA
}}\\
\textit{{$^c$ RIKEN BNL Research Center, Brookhaven National Laboratory, Upton, NY 11973, USA
}}\\
\textit{{$^d$ Department of Physics, Kyoto University, Kyoto 606-8502, Japan
}}\\
\textit{{$^e$
Center for Gravitational Physics and Quantum Information, \\ Yukawa Institute for Theoretical
 Physics, Kyoto University, Kyoto 606-8502, Japan
}}\\
\textit{{$^f$
Kavli Institute for the Physics and Mathematics of the Universe (WPI),\\
 The University of Tokyo, Kashiwanoha, Kashiwa 277-8583, Japan
}}\\
\textit{{$^g$ Faculty of Science and Technology, Seikei University, Musashino, Tokyo 180-8633, Japan
 }}
\end{center}

\vspace*{.5cm}
\begin{abstract}
The D-term is one of the conserved charges of  hadrons defined as the forward limit of  the gravitational form factor $D(t)$. 
We calculate the nucleon's D-term in a holographic QCD model in which the nucleon is described as a soliton in five dimensions. We show that the form factor $D(t)$ is saturated by the exchanges of infinitely many $0^{++}$ and $2^{++}$ glueballs dual to transverse-traceless metric fluctuations on the Wick rotated AdS$_7$ black hole geometry. We refer to this phenomenon as `glueball dominance', in perfect analogy to the vector meson dominance of the electromagnetic form factors. However, the value at vanishing momentum transfer $D(t=0)$ can be interpreted as due to the exchange of 
pairs of pions  and infinitely many vector and axial-vector mesons 
without any reference to glueballs. We find that the D-term is slightly negative as a result of a cancellation between the isovector and isoscalar meson contributions.    


\end{abstract}

\newpage



\date{\today}
%
%
%

\setcounter{section}{+0}
\setcounter{subsection}{+0}

\newpage

%
%
\section{Introduction}
The nucleons (protons and neutrons) have a number of  conserved charges. They have mass of about 1 GeV associated with translational invariance. They have spin  1/2 associated with rotational invariance. They have electric charges and magnetic moments from electromagnetic gauge invariance. They have a baryon number $+1$ associated with the global U(1) symmetry. There are also approximately conserved charges related to isospin (flavor) symmetry. All these charges are well known and very accurately measured, and their values are well documented in textbooks and literature \cite{Zyla:2020zbs}. There is, however, one exactly conserved charge, the so-called D-term, whose  value is presently unknown. The D-term is the forward limit of one of the gravitational form factors $D(t=-k^2)$ defined through the off-forward hadronic  matrix element of the QCD energy momentum tensor $\langle p'|T_{\mu\nu}|p\rangle$ ($k=p'-p$), see \cite{Polyakov:2018zvc} for a recent review.  It was first discovered in the 60's \cite{Kobzarev:1962wt,Pagels:1966zza} but has kept a remarkably low-profile status known only to a subset of theorists in the nuclear physics community. The situation has changed drastically in the past several years, mainly fueled by the anticipation of the future Electron-Ion Collider (EIC) experiments  \cite{AbdulKhalek:2021gbh,Proceedings:2020eah,Anderle:2021wcy} dedicated to the study of the nucleon structure.  
In particular, the first attempt to extract the D-term from experimental data has been made  \cite{Burkert:2018bqq}.

The value of the nucleon D-term has long remained unknown because there is no known way to directly measure it. This would require a controlled experimental setup to scatter a nucleon from gravitons, which is possible only in thought experiments. Yet,  there are indirect ways to measure the D-term in  lepton-nucleon scattering. The price to pay, however, is that one can only access the quark and gluon components separately from  different experiments 
\beq
D=D_u+D_d+D_s+D_g+\cdots,
\eeq
where $D_u$ is from the up-quark part of the energy momentum tensor, $D_g$ is from the gluon part, etc. Each component depends on the renormalization scale, and  only the sum is renormalization-group invariant. The light-quark contribution $D_{u,d}$ can be in principle  accessed in Deeply Virtual Compton Scattering (DVCS) by separately measuring the real and imaginary parts of the Compton form factors  \cite{Teryaev:2005uj,Anikin:2007yh}. 
The gluon contribution $D_g$, on the other hand, can be accessed in near-threshold photo- and lepto-production of heavy quarkonia such as $J/\psi$ and $\Upsilon$ 
\cite{Hatta:2018ina,Hatta:2019lxo,Boussarie:2020vmu,Hatta:2021can,Guo:2021ibg,Kharzeev:2021qkd,Wang:2021dis,Mamo:2022eui} (see however, \cite{Mamo:2019mka,Du:2020bqj,Sun:2021gmi,Mamo:2021tzd,Sun:2021pyw}).  Similarly,  the strangeness  contribution $D_s$ can be probed in near-threshold $\phi$-meson lepto-production  \cite{Hatta:2021can}. Alternatively, these components can also be calculated in lattice QCD simulations, see for example  \cite{Pefkou:2021fni} and references therein.

Although the magnitude and even the sign of the D-term are unknown, a general expectation is that it is negative. This is based on an  analogy to the mechanical stability analysis of classical systems. Being the spatial component of the energy momentum tensor, after Fourier transforming to the coordinate space the D-term form factor may be associated with the `shear' and `pressure' 
\beq
T_{ij}= \left(\frac{r_ir_j}{r^2}-\frac{\delta_{ij}}{3}\right)s(r) + \delta_{ij} p(r) , \label{shear}
\eeq
of a classical spherical system \cite{Polyakov:2002yz,Polyakov:2018zvc}. A positive D-term would imply overall positive outward force, making the system unstable. We however note that there is no field-theoretical proof of the connection between the negativity of the D-term and the stability of hadronic bound states. Besides, the term `pressure' should not be taken literally in its  usual sense in thermodynamics. 

In this paper, we calculate the (total) nucleon D-term in the chiral limit using gauge/string duality based on a  holographic QCD model proposed in \cite{Sakai:2004cn,Sakai:2005yt}. The model is `top-down',  meaning that it has been directly derived from string theory, and is 
realized in a D4/D8 brane configuration in type IIA superstring  theory.
 As such, it does not rely on {\it ad hoc} assumptions and allows for a systematic truncation and/or inclusion of various higher order corrections (stringy effects, 1/$N_c$ corrections, etc.). In this model, the baryons are realized as solitons in a five-dimensional gauge theory \cite{Sakai:2004cn,Nawa:2006gv,Hong:2007kx,Hata:2007mb,Hong:2007ay}, and various properties including electromagnetic form factors \cite{Hong:2007dq,Hashimoto:2008zw,Kim:2008pw} have been investigated using this description. As a matter of fact, holographic approaches are particularly suited for the study of the gravitational form factors   because one can literally exchange gravitons, albeit in extra dimensions. 
   Indeed, there have been previous 
  attempts to compute the D-term in `bottom-up' holographic models  \cite{Abidin:2009hr,Mamo:2019mka,Mamo:2021krl}. However, the outcomes of these studies are rather mixed:  Ref.~\cite{Abidin:2009hr} found that the D-term was simply zero.  Ref.~\cite{Mamo:2019mka} argued that the $D$-form factor was proportional to the $A$-form factor (defined in (\ref{gff}) below), but the proportionality constant remained undetermined. The latter work has been  recently revisited in \cite{Mamo:2021krl,Mamo:2022eui} where  the proportionality constant was found to be subleading in the $1/N_c$ expansion. There is also a work based on an AdS/QCD-inspired quark-diquark  model \cite{Chakrabarti:2020kdc}, but holography was not used in  actual calculations. 
In view of this, it is worthwhile to see what  top-down holographic models  have to say about the D-term. Our model is sophisticated enough to accommodate infinite towers of meson, baryon and  glueball resonances. We shall be particularly interested in how these degrees of freedom contribute to  the D-term. 
 
Our calculation bears some resemblance to  those in the chiral soliton model and the Skyrme model  \cite{Goeke:2007fp,Cebulla:2007ei,Perevalova:2016dln,Kim:2020lrs}. This is so because a baryon in our model is described as a soliton (an instanton) in five-dimensions, similarly to the Skyrmion in four dimensions. In fact, it is known that one can derive the Skyrme model from our model \cite{Sakai:2004cn,Sakai:2005yt,Nawa:2006gv}.   

In the next section, we give a brief review of the gravitational form factors in QCD. In Section 3, we introduce our model and take a first look at the energy momentum tensor in this model. In Section 4, we calculate the D-term in the `classical' approximation by Fourier transforming the soliton energy momentum tensor. Then in Section 5, we discuss the form factor $D(t)$ using holographic renormalization and establish its connection  to scalar and tensor glueballs in QCD. Finally, in Section 6 we conclude with physics interpretations and future perspectives.

\section{Gravitational form factors}

In this section, we quickly introduce  the nucleon gravitational form factors and set up our notations. More details can be found in a recent review \cite{Polyakov:2018zvc}. In accordance with the string theory literature, we use the `mostly plus' metric $\eta^{\mu\nu}=(-1,1,1,1)$.
The QCD energy momentum tensor then takes the from 
\beq
T^{\mu\nu} =
F_a^{\mu\alpha}F^{\nu}_{a\alpha} - \frac{\eta^{\mu\nu}}{4}F_a^{\alpha\beta}F^a_{\alpha\beta} +\bar{\psi}\gamma^{(\mu}D^{\nu)} \psi,
\eeq
where $a=1,2,..,N_c^2-1$ and $A^{(\mu}B^{\nu)}\equiv \frac{A^\mu B^\nu+A^\nu B^\mu}{2}$ denotes symmetrization. $\bar{\psi}=\psi^\dagger \beta = i\psi^\dagger \gamma^0$ and the gamma matrices satisfy the Dirac algebra $\gamma^\mu\gamma^\nu+\gamma^\nu\gamma^\mu=2\eta^{\mu\nu}$. 
The nucleon gravitational form factors are defined by the off-forward matrix element of the energy momentum tensor \cite{Kobzarev:1962wt,Pagels:1966zza}
\beq
\langle p'|T^{\mu\nu}|p\rangle = \bar{u}(p')\left[i\gamma^{(\mu} P^{\nu)} A(t) +\frac{iP^{(\mu}\sigma^{\nu)}_{\ \rho} k^\rho}{2M}B(t) + \frac{k^\mu k^\nu-\eta^{\mu\nu} k^2}{4M}D(t)\right]u(p), \label{gff}
\eeq
where $k=p'-p$, $t=-k^2=k_0^2-\vec{k}^2$ and $P^\mu=\frac{p^\mu+p'^\mu}{2}$. $M$ is the nucleon mass. The nucleon spinors are normalized as $\bar{u}(p)u(p)=2M$.  (\ref{gff}) is the most general parameterization given that $T^{\mu\nu}$ is symmetric and conserved $\partial_\mu T^{\mu\nu}\sim k_\mu T^{\mu\nu}=0$. Energy conservation also implies that the three form factors $A,B,D$ are renormalization group invariant. Their values at $t=0$ are of particular interest. It is known that  $A(t=0)=1$ from momentum conservation and $B(t=0)=0$ from angular momentum conservation. However, the value $D(t=0)$ is not constrained by any symmetry, and  is currently unknown.  Our main goal of this paper is to study $D(t)$ in the holographic QCD model proposed in \cite{Sakai:2004cn,Sakai:2005yt}.

We shall be working in the Breit frame where $\vec{p}{\,'}=\vec{k}/2 = -\vec{p}$, so that $\vec{P}=0$ and $k^0=p'^0-p^0=0$. In this frame,
$\bar{u}(p')u(p)= 2p^0\delta_{s's}=2\sqrt{M^2+\vec{k}^2/4}\,\delta_{s's}$,
where $s',s=\pm\frac12$ denote the spin states of the nucleon,
and
\beq
\frac{\langle p'|T^{00}|p\rangle}{2p^0}&=&
\left(AM -\frac{B\vec{k}^2}{4M} + \frac{D\vec{k}^2}{4M}\right)\delta_{s's}, \nn 
\frac{\langle p'|T^{0i}|p\rangle}{2p^0}&=& \frac{A+B}{4}\epsilon_{ijk}(-ik^j)\sigma^k_{s's}, \label{br}\\ 
\frac{\langle p'|T^{ij}|p\rangle}{2p^0}&=& \frac{D}{4M}(k^i k^j-\vec{k}^2\delta^{ij})\delta_{s's}. \qquad (i,j=1,2,3) \nonumber
\eeq
Therefore, the D-term can be obtained by  reading off the coefficient of $k^ik^j$ in $\langle T^{ij}\rangle$.

For a later discussion, following \cite{Hatta:2018ina}, let us also introduce the `transverse-traceless' (TT) part $T^{\mu\nu}_{\rm TT}$ of the energy momentum tensor. It is defined as a part of  $T_{\mu\nu}$ that satisfies the conditions $\partial_\mu T^{\mu\nu}_{\rm TT}=(T_{\rm TT})^\mu_\mu=0$. Explicitly, 
\beq
T_{\rm TT}^{\mu\nu}= T^{\mu\nu} +\frac{1}{3}\left(\frac{\partial^\mu \partial^\nu}{\partial^2} -\eta^{\mu\nu}\right)T^\mu_\mu.
\eeq
Eq.~(\ref{gff}) can then be rewritten as 
\beq
\langle p'|T^{\mu\nu}|p\rangle &=& \bar{u}(p')\left[i\gamma^{(\mu} P^{\nu)} A +\frac{iP^{(\mu}\sigma^{\nu)}_{\ \rho} k^\rho}{2M}B - \frac{M}{3} \left(k^\mu k^\nu -\eta^{\mu\nu}k^2\right)\left(\frac{A}{k^2}-\frac{B}{4M^2}\right)\right]u(p) \nn
&&+\frac{M}{3} \left(k^\mu k^\nu-\eta^{\mu\nu}k^2\right)\left(\frac{A}{k^2} - \frac{B}{4M^2}+\frac{3D}{4M^2} \right)\bar{u}(p')u(p).  \label{can}
\eeq
The first line on the right hand side is the matrix element of $T_{\rm TT}^{\mu\nu}$ and the second line is from the trace part. Note that the structure $k^\mu k^\nu-k^2\eta^{\mu\nu}$ characteristic of the D-term is now present in both parts.

Before leaving this section, for the convenience of the reader, we  note the  definition of the gravitational form factors in the `mostly minus' metric $\eta^{\mu\nu}=(1,-1,-1,-1)$ which is used in most QCD literature. 
 The gamma matrices in the two conventions are related as 
\beq
\gamma^\mu_{\rm mostly \, minus} = i\gamma^\mu_{\rm mostly \, plus}.
\eeq
With the QCD energy momentum tensor in this metric 
\beq
T^{\mu\nu} = 
 -F^{\mu\alpha}F^{\nu}_{\ \alpha} + \frac{\eta^{\mu\nu}}{4}F^{\alpha\beta}F_{\alpha\beta} + \bar{\psi}i\gamma^{(\mu}D^{\nu)} \psi,
\eeq
($\bar{\psi}=\psi^\dagger\gamma^0$), we now write  
\beq
\langle p'|T^{\mu\nu}|p\rangle = \bar{u}(p')\left[\gamma^{(\mu} P^{\nu)} A(t) +\frac{iP^{(\mu}\sigma^{\nu)}_{\ \rho} k^\rho}{2M}B(t) + \frac{k^\mu k^\nu-\eta^{\mu\nu} k^2}{4M}D(t)\right]u(p),
\eeq
where $t=k^2=k_0^2-\vec{k}^2$.

\section{Nucleon in holographic QCD}

The two-flavor ($N_f=2$) meson-baryon sector of our model \cite{Sakai:2004cn,Sakai:2005yt} is defined by a U(2) gauge theory in a curved five-dimensional spacetime $(x^{\mu=0,1,2,3},z)$ supplemented with the Chern-Simons (CS) term  
\beq
S = -\kappa \int d^4xdz \, {\rm tr} \left[\frac{1}{2}h(z)F_{\mu\nu}^2+k(z)F_{\mu z}^2\right]-\frac{\kappa}{2} \int d^4xdz  \left(\frac{1}{2}h(z)\widehat{F}_{\mu\nu}^2+k(z)\widehat{F}_{\mu z}^2\right)+S_{CS}, \label{action}
\eeq
where 
\beq
h(z)=(1+z^2)^{-1/3}, \qquad k(z)=1+z^2, \qquad  (-\infty<z<\infty)
\eeq
are the warp factors along the fifth dimension. $F_{\mu\nu}=F_{\mu\nu}^a \frac{\tau^a}{2}$, $F_{\mu z}=F_{\mu z}^a \frac{\tau^a}{2}$ are the SU(2) field strength tensors ($\tau^{a=1,2,3}$ being the Pauli matrices normalized as ${\rm tr}[\tau^a\tau^b]=2\delta^{ab}$), and $\widehat{F}_{\mu\nu}$ is the U(1) field strength tensor. The Lorentz indices of these gauge fields are raised and lowered  by the flat five-dimensional metric $(-1,1,1,1,1)$.  The Chern-Simons term $S_{CS}$ prescribes the interaction between the SU(2) and U(1) fields, but its explicit form is not needed for the present discussion. 
The parameter  
\beq
\kappa=\frac{\lambda N_c}{216\pi^3},\label{kappa1}
\eeq
is proportional to the number of colors $N_c$ and the `t Hooft coupling  $\lambda=g^2N_c$. 
All dimensionful  scales have been made dimensionless by appropriately  rescaling by  the model's only mass parameter $M_{KK}$ (e.g., $x^\mu\to x'^\mu=x^\mu M_{KK}$). These parameters was determined in \cite{Sakai:2004cn,Sakai:2005yt} by fitting the $\rho$-meson mass  and the pion decay constant 
\beq
M_{KK}=949\, {\rm MeV}, \qquad \lambda=16.63. \quad  (\kappa=0.00745) \label{mkk}
\eeq
 
Mesons with isospin quantum numbers $\pi,\rho,a_1,\cdots$ are described by the fluctuations of the SU(2) gauge field $F_{\mu\nu}$. We consider the chiral limit, so the pions are massless. Iso-singlet mesons $\omega, \eta',\cdots$ are described by the U(1) field.  A baryon is realized by a static (independent of $t=x^0$), soliton-like configuration of $F_{\mu\nu}$ which satisfies the equation of motion of the five-dimensional gauge theory (\ref{action}) \cite{Sakai:2004cn,Hata:2007mb}. This is charged under the U(1) gauge field $\widehat{A}_{\mu}$ through the CS term, and the charge is identified with the baryon number. At strong coupling $\lambda \gg 1$ and in the small-$z$ region where the metric is  approximately flat $h(z)\approx k(z)\approx 1$, the solution is simply given by the BPST instanton \cite{Belavin:1975fg} in  four-dimensional Euclidean space $(x^{i=1,2,3},z)$ \cite{Hata:2007mb}
\beq
&&A_i=
\frac{(z-Z)\tau_i+(x_m-X_m) \epsilon_{min}\tau_n}{\xi^2+\rho^2}, \qquad  
A_z=\frac{-(\vec{x}-\vec{X})\cdot \vec{\tau}}{\xi^2+\rho^2},  \nn
&& F_{ij}=\partial_iA_j-\partial_jA_i+i[A_i,A_j]=\frac{2\rho^2}{(\xi^2+\rho^2)^2}\epsilon_{ija}\tau^a, \qquad F_{iz}=-\frac{2\rho^2}{(\xi^2+\rho^2)^2}\tau_i, \label{instanton} \\
&&\widehat{A}_0 = 
\frac{27\pi}{\lambda } \frac{2\rho^2+\xi^2}{(\rho^2+\xi^2)^2}, \nonumber
\eeq
 where $\xi^2\equiv (\vec{x}-\vec{X})^2+(z-Z)^2$. $\rho$ is the instanton `size' and  $(\vec{X},Z)$ denotes the `center' of the instanton. When the soliton is quantized, these parameters, together with the `orientation' of the instanton in the flavor SU(2) space  $F_{\mu\nu}\to VF_{\mu\nu}V^{-1}$, are promoted to time-dependent operators (for example, $\rho \to \hat{\rho}(t)$), a procedure known as the collective coordinate quantization. While this has been done in previous applications of the model \cite{Hata:2007mb,Hashimoto:2008zw}, in this work we shall   treat them as c-numbers, leaving their quantum treatment for future work. This means that we eventually set $\vec{X}=0$ (without loss of generality) and employ the value $Z=0$ which minimizes the soliton potential in the $Z$-direction \cite{Hata:2007mb}.  It should be kept in mind that, by neglecting quantization, we are effectively treating the nucleon as a scalar particle because the quantization of the SU(2) orientation is what makes the soliton a spin-1/2 particle.  The $D$-form factor exists also for scalar hadrons, but the $B$-form factor (see (\ref{gff})) does not.  In order to compute the latter, soliton quantization is crucial.

 When $|z|\gtrsim 1$, the flat space approximation $h\approx k\approx 1$  breaks down.  Exact analytical solutions in this region are no longer available.  However, when $|z|\gg 1$,  the following approximate solution has been constructed  \cite{Hashimoto:2008zw} ($\partial_i \equiv \partial/\partial x^i$) 
\begin{eqnarray}
&& \widehat{A}_0\approx  \frac{-108\pi^3}{\lambda}  G,
\nn
&&A^g_i\approx 2\pi^2\rho^2 \tau^a(\epsilon_{ial}\partial_l + \delta_{ia}\partial_Z)G, \nn
&& A^g_z\approx 2\pi^2\rho^2\tau^a \partial_a H,
 \label{sch1} \\
&& 
F^g_{iz} \approx 2\pi^2\rho^2\tau^a\Bigl[ \partial_i\partial_a H -\delta_{ia}\partial_z\partial_ZG -\epsilon_{iak}\partial_k \partial_zG  \nn
&& \qquad \qquad -4 
\pi^2\rho^2 \left(\delta_{ia}\partial_k G\partial_kH -\partial_a G\partial_iH-\epsilon_{iac}\partial_ZG\partial_cH\right)\Bigr], \nn
&&F^g_{ij}\approx    2\pi^2\rho^2\tau^a \Bigl[\left(\epsilon_{jal}\partial_i \partial_l -\epsilon_{ial}\partial_l \partial_j -\delta_{ia}\partial_j \partial_Z +\delta_{ja}\partial_i \partial_Z\right) G \nn 
&& \qquad \qquad -4\pi^2\rho^2\left(\epsilon_{aij}(\partial_Z G)^2 + \epsilon_{ijl}\partial_aG \partial_l G +\delta_{ia}\partial_jG \partial_Z G -\delta_{ja}\partial_iG\partial_Z G\right) \Bigr],
\nonumber
\end{eqnarray}
in the so-called singular gauge 
\beq
A \to A^g= g^{-1}A g -ig^{-1}d g, \qquad g= \frac{z-Z -i(\vec{x}-\vec{X})\cdot \vec{\tau}}{\xi}.
\eeq
In \cite{Hashimoto:2008zw}, the non-Abelian commutator terms in $F^g_{iz}$ and $F^g_{ij}$ have been dropped since they are negligible when $|z|\gg 1$. Here we have restored them for a later purpose.   
The Green functions $G(|\vec{x}-\vec{X}|,z,Z)$ and $H(|\vec{x}-\vec{X}|,z,Z)$ satisfy  ($\nabla_x^2\equiv \partial_i\partial_i$)
\begin{equation}
\begin{split}
& h(z)\nabla_x^2G+\partial_z (k(z)\partial_z G)=\delta(z-Z)\delta(\vec{x}-\vec{X}) ,\\
& \nabla_x^2 H + \partial_z\Bigl(h^{-1}(z)\partial_z(k(z)H)\Bigr)=\frac{1}{k(z)}\delta(z-Z) \delta(\vec{x}-\vec{X}). \label{sch2}
\end{split}
\end{equation}

In order to compute the D-term, or more generally the gravitational form factors, it is desirable to have an approximate solution which smoothly interpolates the above solutions in the two limits $|z|\ll 1$ and $|z|\gg 1$. Such a solution can be readily found  for the U(1) part. We start with the small-$z$ region and consider the following equation of motion for $\widehat{A}_0$ in the SU(2) instanton background
\beq
h(z)\nabla_x^2 \widehat{A}_0+ \partial_z (k(z)\partial_z\widehat{A}_0)=-\frac{648\pi}{\lambda} \frac{\rho^4}{(\xi^2+\rho^2)^4}, \label{improve}
\eeq
or in the Fourier space, 
\beq
-\vec{k}^2h(z) \widehat{A}_0 +  \partial_z (k(z)\partial_z\widehat{A}_0) 
=- \frac{27\pi^3\rho^4e^{-|\vec{k}|\sqrt{\rho^2+z^2}}}{\lambda (\rho^2+z^2)^{5/2}}\left(3+3|\vec{k}|\sqrt{\rho^2+z^2}+\vec{k}^2(\rho^2+z^2)\right). \label{fourierA0}
\eeq
In the flat space approximation $h(z)\approx k(z)\approx 1$, the solution regular at $\xi=0$ is given by  (\ref{instanton}). 
The boundary conditions are such that $\widehat{A}_0(z\to \infty)=0$ and  the solution is smooth at $z=0$, namely, $\partial_z \widehat{A}_0(z)|_{z=0}=0$. 
When $|z|\gg 1$, the right hand side of (\ref{improve}) is negligible, and the equation becomes identical to that for $G$ in (\ref{sch2}). Therefore, the solution of (\ref{improve}) with the said boundary conditions smoothly interpolates the solutions at $|z|\ll 1$ and $|z|\gg 1$. We will use it as an approximate solution in the whole region $0<|z|<\infty$. 

The situation is more complicated for the SU(2) part. The asymptotic solution (\ref{sch1}) with (\ref{sch2}) has been obtained by neglecting the nonlinear terms in the Yang-Mills equation. In the small-instanton regime $\rho\ll 1$, or equivalently,  the strong coupling regime $\lambda\gg 1$ (note the correspondence $\rho\sim 1/\sqrt{\lambda}$  \cite{Hata:2007mb}), the large-$z$ and small-$z$ solutions have an overlapping region of validity $\rho \ll  z \ll 1$ where they can be smoothly matched  \cite{Hashimoto:2008zw}. 
We extrapolate $G$ and $H$ to small-$z$   by solving  
  (\ref{sch1}) with the instanton configuration (\ref{instanton}) rotated to the singular gauge $A_{\mu}\to A_{\mu}^g$ and substituted into the left hand side. This gives 
\beq
G(x,z)=H(x,z)=\frac{1}{4\pi^2\rho^2}\ln \frac{\xi^2}{\xi^2+\rho^2}, \qquad (|z|\lesssim 1)  \label{kfourier}
\eeq
with 
\beq
(\nabla_x^2+\partial_z^2)\begin{pmatrix} G \\ H \end{pmatrix}
= \frac{\rho^2}{\pi^2\xi^2(\xi^2+\rho^2)^2}. \label{alter}
\eeq
In the $\rho\to 0$ limit, $G,H$ reduce to the flat space Green function $\frac{-1}{4\pi^2\xi^2}$ in four dimensions. Regarding the right hand side of (\ref{alter}) as a regularized form of  $\delta(\vec{x}-\vec{X})\delta(z-Z)$, we are led to consider the following equations
\begin{equation}
\begin{split}
& h(z)\nabla_x^2G+\partial_z (k(z)\partial_z G)= \frac{\rho^2}{\pi^2\xi^2(\xi^2+\rho^2)^2}, \label{newG}
\\
& \nabla_x^2 H + \partial_z\Bigl(h^{-1}(z)\partial_z(k(z)H)\Bigr)=\frac{h(z)}{k(z)} \frac{\rho^2}{\pi^2\xi^2(\xi^2+\rho^2)^2},
\end{split}
\end{equation}
instead of (\ref{sch2}).\footnote{Differently from (\ref{sch2}), we have introduced  $h(z)$ on the right hand side of the equation for $H$ for a technical reason to be explained in the next section. The choice of the inhomogeneous term is somewhat arbitrary in our  construction of an approximate solution.}
Eq.~(\ref{newG}) is to be solved 
with boundary conditions $G,H\to 0$ as $|z|\to \infty$ and $\partial_z G|_{z=0}=\partial_z H|_{z=0}=0$. At large-$z$, the right hand side is negligible, and the equation reduces to (\ref{sch2}). At small-$z$, $G$ and $H$ smoothly connect to the instanton solution by construction. Therefore, we can use (\ref{sch1}) with  (\ref{newG}) as an approximate solution in the whole range $0<|z|<\infty $ even when $\rho$ is order unity, provided the non-Abelian commutator terms in $F_{ij}$ and $F_{iz}$ in (\ref{sch1}) are kept. 
To go beyond this approximation, one has to numerically solve the Yang-Mills equation in the  curved background as was done in   \cite{Bolognesi:2013nja,Rozali:2013fna,Suganuma:2020jng}.

Let us now take a first look at the energy momentum tensor in this model. We can  write down  the following `classical' energy momentum tensor in four-dimensions
\beq
T^{cl}_{\mu\nu}(x)&=& 2\kappa\int^\infty_{-\infty} dz\,  {\rm tr}\left[ h(z)F_{\mu\rho}F_{\nu}^{\ \rho}+ k(z) F_{\mu z}F_{\nu z} -\frac{\eta_{\mu\nu}}{2}\left(\frac{h(z)}{2}F_{\alpha\beta}^2 +k(z)F_{\mu z}^2 \right)\right] \nn 
&& +\kappa \int^\infty_{-\infty} dz  \left[ h(z)\widehat{F}_{\mu\rho}\widehat{F}_{\nu}^{\ \rho}+ k(z) \widehat{F}_{\mu z}\widehat{F}_{\nu  z} -\frac{\eta_{\mu\nu}}{2}\left(\frac{h(z)}{2}\widehat{F}_{\alpha\beta}^2 +k(z)\widehat{F}_{\mu z}^2 \right)\right],
\label{return}
\eeq
by varying the action (\ref{action}) with respect to the flat metric $\eta^{\mu\nu}$ \cite{Karch:2008uy}. 
The Chern-Simons term does not contribute since it is independent of the metric. When evaluated  on-shell,  (\ref{return})  is conserved 
\beq
\partial^\mu T_{\mu\nu}^{cl}=0. \label{conserve}
\eeq
We have explicitly verified (\ref{conserve})  using the equation of motion derived in  \cite{Hashimoto:2008zw}. 

We expect that (\ref{return}) is a reasonable approximation to the full result, when the baryon is treated as heavy (at least parametrically)  as in large-$N_c$ QCD. (In Section 5, we shall discuss the corrections to this formula due to glueballs.) 
In particular, 
the nucleon mass can be calculated as
\beq
M=\int d^3x T^{cl}_{00}(\vec{x}) &=& \kappa \int d^3x dz {\rm tr}\left[\frac{h(z)}{2}F_{ij}^2 +k(z)F_{i z}^2\right] \nn 
&& +\frac{\kappa}{2} \int d^3x dz \left[h(z) (\partial_i \widehat{A}_0)^2+k(z)(\partial_z \widehat{A}_0)^2 \right], \label{saka}
\eeq
where we have already substituted the classical  configuration $A_0=\widehat{A}_i=\widehat{A}_z=0$. Eq.~(\ref{saka})  is consistent with Eq.~(3.18) of \cite{Hata:2007mb} after using the equation of motion. The latter expression is simply the minus of the on-shell action $\int dt M=-S$ (this time including the Chern-Simons term), which is appropriate for a classical, heavy particle at rest. 
For the instanton solution, the integrals in (\ref{saka}) can be explicitly evaluated and lead to the  structure  \cite{Hata:2007mb}
\beq
M= 8\pi^2\kappa\left(1+{\cal O}(\rho^2)+{\cal O}\left(\frac{1}{\lambda^2\rho^2}\right)\right). \label{stable}
\eeq
(Remember the mass is measured in units of $M_{KK}$.)
The leading term $8\pi^2\kappa \sim {\cal O}(\lambda N_c)$ comes from the SU(2) part. The subleading terms $\rho^2$ and $1/\rho^2$ come from  the SU(2) and U(1) fields, respectively. The SU(2) fields tend  to shrink the instanton size  $\rho\to 0$, while the U(1) field tends to expand the  size $\rho\to \infty$, making the system unstable. As a compromise, the minimum energy is achieved when $\rho\sim {\cal O}(1/\sqrt{\lambda})$.  The situation is entirely analogous to the Skyrme model where the baryon (realized as a solitonic configuration of pions) is stabilized by introducing the $\omega$-meson \cite{Adkins:1983nw}.  
The numerical value of $\rho$ has been fixed in this way in the instanton approximation  \cite{Hata:2007mb}
\beq
\rho=\sqrt{\frac{27\pi}{\lambda}\sqrt{\frac{6}{5}}}\approx 2.36, \qquad \frac{M}{M_{KK}}= \frac{\lambda N_c}{27\pi} + \sqrt{\frac{2}{15}}N_c \approx 1.68. \label{parameter2}
\eeq
Together with (\ref{mkk}), this gives $M\sim 1.5$ GeV which is larger than the observed value. However, the value of $M$ is subject to changes  after the collective coordinate quantization \cite{Hata:2007mb}.

Returning to (\ref{return}), our main interest in this paper is the spatial $i,j=1,2,3$ components
\beq
T^{cl}_{ij}(\vec{x}) &=& 2\kappa \int^\infty_{-\infty} dz\,  {\rm tr}\left[ h(z)F_{il}F_{jl}+ k(z) F_{i z}F_{jz} -\frac{\delta_{ij}}{2}\left(\frac{h(z)}{2}F_{lm}^2 +k(z)F_{l z}^2 \right)\right] \nn
&& +\kappa \int^\infty_{-\infty} dz  \left[- h(z)\partial_i \widehat{A}_0 \partial_j \widehat{A}_0  +\frac{\delta_{ij}}{2}\left( h(z)(\partial_i \widehat{A}_0)^2 + k(z) (\partial_z \widehat{A}_0)^2 \right) \right] .
\label{zint}
\eeq 
Classically, one might expect that the form factor $D(|\vec{k}|)$ could be calculated by simply inserting the above solutions into (\ref{zint}) and Fourier transforming to momentum space $\vec{x}\to \vec{k}$. (Below we shall often use the notation $D(|\vec{k}|)$  instead of $D(t)$.) Since $T^{cl}_{ij}$ is conserved, it must have the structure 
\beq
  T_{ij}^{cl}(\vec{k}) = (k_ik_j-\delta_{ij}\vec{k}^2)\frac{D(|\vec{k}|)}{4M}.  \label{dterm}
 \eeq
 However, it turns out that this naive approach is valid only at $\vec{k}=0$.  
In the next section, we shall numerically evaluate the D-term $D(0)$  in this way. A more general analysis valid for arbitrary values of  $|\vec{k}|$ will be presented in Section 5.

\section{`Classical' calculation of the D-term}

In this section, we calculate the D-term `classically', by Fourier transforming the naive energy momentum tensor (\ref{zint}) with the approximate solutions $\widehat{A}_0$, $F_{ij}$ and $F_{iz}$ constructed in the previous section. This is analogous to what has been done in the Skyrme model or  chiral soliton models. For a reason to be  clarified in the next section, the calculation in the present section is  valid only at vanishing momentum transfer $|\vec{k}|=0$.  
 When $|\vec{k}|\neq 0$, one must carry out a fully holographic calculation which  will be discussed in the next section.  
  We shall calculate the contributions from the U(1) and SU(2) fields separately 
\beq
D(0)=D^{\rm U(1)}(0)+D^{\rm SU(2)}(0).
\eeq
In the remainder of this section, only three-momenta $\vec{k}$, $\vec{q}$ appear. We thus  write  $|\vec{k}|\equiv k$, $|\vec{q}|\equiv q$ below to simplify the notation.

\subsection{U(1) part}

In Section 3, we have constructed an approximate solution for the U(1) gauge potential  $\widehat{A}_0(x,z)$ which can be used in the entire range $-\infty <z<\infty$. This is obtained by numerically solving  (\ref{fourierA0}) with the boundary conditions $\widehat{A}_0(k,z=\infty)=0$ and $\partial_z \widehat{A}(k,z)|_{z=0}=0$.  The next step is to substitute this solution into (\ref{zint}). 
\beq
T_{ij}^{\rm U(1)}
&=& \kappa \int^\infty_{-\infty} dz  \int d^3x e^{i\vec{k}\cdot \vec{x}}\left[-h(z)\partial_i \widehat{A}_0 \partial_j\widehat{A}_0  + \frac{\delta_{ij}}{2}\left(h(z)(\partial_l\widehat{A}_0)^2+k(z)(\partial_z\widehat{A}_0)^2\right) \right] 
\nn
&= &
\kappa \int^\infty_{-\infty} dz  \int \frac{d^3q}{(2\pi)^3} \Biggl[ h(z) q_i (k_j-q_j)\widehat{A}_0(q,z)\widehat{A}_0(|\vec{k}-\vec{q}|,z) \nn
&& \qquad \qquad \qquad -\frac{\delta_{ij}}{2}\left(h(z)\vec{q}\cdot(\vec{k}-\vec{q})\widehat{A}_0(q,z)+\partial_z(k(z)\partial_z\widehat{A}_0(q,z))\right)\widehat{A}_0(|\vec{k}-\vec{q}|,z)\Biggr] \nn 
&\equiv & \kappa \int^\infty_{-\infty} dz  ( P_1(k,z)k_i k_j + Q_1(k,z) \delta_{ij}k^2). \label{nc}
\eeq
In the second equality we integrated by parts in $z$. The coefficients $P_1$ and $Q_1$ can be calculated as follows  
\beq
P_1(k,z)&=& \frac{h(z)}{2(k^2)^2} \int \frac{d^3q}{(2\pi)^3}(2\vec{q}\cdot \vec{k} k^2 -3(\vec{q}\cdot \vec{k})^2+q^2k^2)\widehat{A}_0(q,z)\widehat{A}_0(|\vec{k}-\vec{q}|,z)  \label{u11} \nn 
&=&  \frac{h(z)}{8\pi^2k^2} \int_{-1}^1 d\cos \theta \nn 
&& \times \int_0^\infty q^3 dq
(2k\cos \theta-3q\cos^2\theta+q)\widehat{A}_0(q,z)\widehat{A}_0(\sqrt{k^2+q^2-2kq\cos\theta},z).   \label{p1}
\eeq
\beq
Q_1(k,z)= \frac{1}{2k^2}\int \frac{d^3q}{(2\pi)^3}\left[h(z)\left(\frac{(\vec{q}\cdot \vec{k})^2}{k^2}-\vec{q}\cdot \vec{k}\right)\widehat{A}_0(q,z)-\partial_z(k(z)\partial_z\widehat{A}_0(q,z))\right]\widehat{A}_0(|\vec{k}-\vec{q}|,z).
\eeq
 Despite the singular prefactor $1/k^2$, $P_1(k=0,z)$ is  finite as one can see by expanding the integrand in $k$ and performing the angular integral. On the other hand,  $Q_1(k=0,z)$ is actually divergent. If the bulk equation of motion is solved exactly, this divergence should be  canceled by the contribution from the SU(2) field. Moreover, after the cancellation the coefficients of $k_ik_j$ and   $-\delta_{ij}k^2$ must agree exactly due to the conservation law (cf. (\ref{dterm})).  Indeed, after some manipulations (including the addition of total derivative terms), one can show that
 \beq
 Q_1(k,z)&=&-P_1(k,z) \\
 && -\frac{1}{(k^2)^2} \int \frac{d^3q}{(2\pi)^3} \vec{k}\cdot(\vec{k}-\vec{q}) \left(-q^2 h(z) \widehat{A}_0(q,z) +\partial_z (k(z)\partial_z \widehat{A}_0(q,z))\right) \widehat{A}_0(|\vec{k}-\vec{q}|,z). \nonumber
 \eeq
 The expression inside the brackets is connected to the SU(2) fields via the equation of motion (\ref{fourierA0}).  In practice, since our solution is  approximate and obtained only numerically, it may be difficult to achieve a precise  cancellation. We thus focus on the coefficient of $k_ik_j$  which is safely calculable in the present approach, and leave a more complete analysis for future work. 

 \subsection{SU(2) part}
 
 We now turn to the SU(2) fields. Let us first point out that the instanton solution, valid in the small-$z$ region, does not give rise to the structure  $k_ik_i$. Indeed, inserting  (\ref{instanton}) into (\ref{zint}), we find
 \beq
 T_{ij}^{\rm instanton}(k) &\sim& 8\kappa \rho^4 \delta_{ij} \int dz \int d^3x e^{ik\cdot x} \frac{h(z)-k(z)}{(\xi^2+\rho^2)^4} \nn
&=& \frac{\pi^2\kappa \rho^4 \delta_{ij}}{3}\int dz \frac{h(z)-k(z)}{(z^2+\rho^2)^{5/2}}\left(3+3k\sqrt{z^2+\rho^2}+k^2(z^2+\rho^2)\right) e^{-k\sqrt{z^2+\rho^2}}, \label{instD}
 \eeq
where the $z$-integral should be cut off around $z\lesssim {\cal O}(1)$.  
Note that the coefficient of $\delta_{ij}$ is finite at $k=0$, meaning that  (\ref{instD}) gives a divergent contribution to $D(k=0)$ since $T_{ij} \sim D(k)\delta_{ij}k^2$. This is  expected  in view of our discussion in the previous subsection. If the equation of motion is solved exactly, the divergence must be canceled by the contributions from the U(1) field  as well as that from the SU(2) field in the  large-$z$ region (where the instanton approximation breaks down). However, since we decided to focus on the coefficient of $k_ik_j$, we do not dwell on  (\ref{instD}) further. 
 
Next, we consider the large-$z$ region $|z|\gg 1$  where the solution is given by (\ref{sch1}) and (\ref{sch2}). Eq.~(\ref{sch2}) can be formally solved by introducing the complete set of eigenfunctions   \cite{Sakai:2004cn,Hashimoto:2008zw}, 
\beq
-\frac{1}{h(z)}\partial_z(k(z)\partial_z \psi_n(z))=m_n^2 \psi_n(z), \qquad (n=1,2,3,\cdots)
\eeq
normalized as  
\beq
\kappa \int dz h(z) \psi_n\psi_m = \delta_{mn},
\eeq
and associated eigenfunctions 
\beq
\phi_0(z)=\frac{1}{\sqrt{\kappa \pi}} \frac{1}{k(z)}, \qquad \phi_n(z)=\frac{1}{m_n}\partial_z \psi_n(z), \qquad (n=1,2,3,\cdots)
\eeq
Using these eigenfunctions, we can write the solution of (\ref{sch2}) as, in momentum space,    
\begin{equation}
\begin{split}
G(k,z,Z)=-\kappa \sum_{n=1}^{\infty}\frac{\psi_n(z)\psi_n(Z)}{k^2+m_n^2},  \\
H(k,z,Z)=-\kappa \sum_{n=0}^{\infty} \frac{\phi_n(z)\phi_n(Z)}{k^2+m_n^2}, \label{mesons}
\end{split}
\end{equation}
where $m_0=0$. 
$\psi_n(z)$ is an even (odd) function in $z$ when $n$ is odd (even). Since $\psi_{2n}(0)=0$ and $\phi_{2n+1}(0)=0$, only odd $n$ values contribute in $G$, and only even $n$ values contribute in  $H$ at $Z=0$.  
Numerically we find, for $n\ge 1$, 
\beq
\frac{m_n}{M_{KK}} = 0.818, \quad 1.2525, \quad 1.695, \quad 2.132, \quad 2.567, \quad 3.001, \quad 3.435,\quad 3.868, \quad 4.300,... \label{masses}
\eeq 
Note that $m_1=0.818M_{KK}\approx 776$ MeV is the $\rho$-meson mass. More generally, $G\sim \psi_{2n-1}$ represents vector mesons ($\rho$-meson excited states) and $H\sim \phi_{2n}$ with $n\ge 1$ represents axial vector mesons ($a_1$-meson excited states). The $n=0$ term $\phi_0$ in $H$ represents the massless pion field. 

To calculate the D-term, we need to extrapolate the above solution to the small-$z$ region. The main complication is the nonlinearity of the SU(2) Yang-Mills equation. We have  argued   in Section 3  that, at least in the strong coupling limit where $1\gg \rho$, the linear approximation is valid up to $z\gtrsim \rho$, and an approximate solution in this regime can be obtained  by replacing (\ref{sch2}) with  (\ref{newG}). The remaining  parametrically small region $\rho >z>0$  does not contribute to the coefficient of $k_ik_j$ as we have seen in (\ref{instD}), so we may further extrapolate this solution down to $z=0$. However, as the coupling is lowered and $\rho$ exceeds unity, the non-Abelian commutator terms in $F_{ij}$ and $F_{iz}$ become important. This is why we have kept them in  (\ref{sch1}). 

Since the full SU(2) energy momentum tensor is local and involves up to four powers of the gauge fields, it is much more advantageous to work in the original coordinate space.  
In (\ref{zint}), there are two terms that can give rise to the structure $k_ik_j$. Noting that $G$ and $H$ depend only on the magnitude $x=|\vec{x}|$ (and $z$), we  write  
\beq
  {\rm tr}[h(z)F_{il}F_{jl} + k(z) F_{iz}F_{jz}] =\frac{x_ix_j}{x^2}X(x,z)+\delta_{ij}Y(x,z).
\eeq
After a rather tedious calculation we find 
\beq
X=2(2\pi^2\rho^2)^2 \bigl[ h(z)({\cal E}^2+{\cal F}^2-{\cal I}^2+2{\cal D} {\cal F}-2{\cal D} {\cal I}) + k(z)(2{\cal A}{\cal B}+{\cal B}^2-{\cal C}^2) \bigr], \label{tedious}
\eeq
with 
\beq
&&{\cal A}= \frac{\partial_x H}{x} -\partial_z\partial_ZG -4\pi^2\rho^2\partial_xG\partial_x H, \nn 
&&{\cal B}= \partial_x^2H-\frac{\partial_x H}{x} + 4\pi^2\rho^2\partial_x G\partial_x H, \nn
&&{\cal C}=\partial_x \partial_z G -4\pi^2\rho^2\partial_Z G \partial_x H, \nn
&&{\cal D}=\frac{2\partial_x G}{x}-4\pi^2\rho^2(\partial_Z G)^2, \label{abc} \\
&&{\cal E}=\partial_x \partial_Z G +4\pi^2\rho^2 \partial_xG \partial_ZG, 
\nn
&&{\cal F}=\partial_x^2G-\frac{\partial_x G}{x}, \nn
&& {\cal I}= -4\pi^2\rho^2(\partial_xG)^2. \nonumber
\eeq
$\partial_Z G$  can be eliminated by using the formula (see Eq.~(2.79) of \cite{Hashimoto:2008zw} and (\ref{newG}))
\beq
 \partial_Z G = -\frac{1}{h(z)}\partial_z(k(z)H), \qquad \partial_z \partial_Z G = \nabla^2_x H - \frac{h(z)}{k(z)} \frac{\rho^2}{\pi^2\xi^2(\xi^2+\rho^2)^2},
\eeq
after which we may set $Z=0$. The first relation was originally derived in the large-$z$, linear regime, but it is  also valid in the small-$z$ regime where $k(z)\approx h(z) \approx 1$, $G\approx H$ and $\partial_Z G \approx - \partial_z G$.  
Given $X$, we can evaluate  the SU(2) contribution to the D-term as 
\beq
D^{\rm SU(2)}(0)=-64\pi \kappa M\lim_{k\to 0} \frac{1}{k^2}\int_0^\infty dz \int_0^\infty dx x^2 j_2(kx) X(x,z), \label{dsu2}
\eeq
where $j_2$ is the spherical Bessel function.

\subsection{Numerical result}

We have solved (\ref{fourierA0}) and (\ref{newG}) numerically with the Neumann boundary conditions at $x=0$ and $z=0$ and the Dirichlet boundary conditions $\widehat{A}_0=G=H=0$ at infinity. Great care is needed in order to obtain a stable solution for $H$ (and especially its $x,z$-derivatives) because of the massless pion pole, the $n=0$ term in  (\ref{mesons}), as well as the pole $1/\xi^2$ on the right hand side of (\ref{newG}). To cope with these, 
we first make the following shift (cf. \eqref{kfourier})
\begin{equation}
  G(x,z) = \tilde{G}(x,z) + \frac{1}{4\pi^2\rho^2} \ln\frac{\xi^2}{\xi^2+\rho^2}, \qquad
  H(x,z) = \tilde{H}(x,z) + \frac{1}{k(z)} \frac{1}{4\pi^2\rho^2} \ln\frac{\xi^2}{\xi^2+\rho^2},
  \label{subtraction}
\end{equation}
to avoid the singular behavior  near the origin.\footnote{This is why we have introduced the factor $h(z)$ on the right hand side of the equation for $H$ in (\ref{newG}). Without this factor, the  subtraction (\ref{subtraction}) induces a discontinuity in the resultant equation at the origin: The limits $x\to 0$ with $z=0$ fixed and  $z\to 0$ with $x=0$ fixed do not agree. }    
We then solve the resultant differential equations for $\tilde{G}(x,z)$ and $\tilde{H}(x,z)$ by  employing the pseudospectral method \cite{boyd2001chebyshev}. In this method, the functions $\tilde{G}(x,z)$, $\tilde{H}(x,z)$ are expanded by tensor products $u_m(x;L_x) u_n(z;L_z)$ ($1\le m,n\le N$) where the basis functions $u_n(y;L) = y^2/(y+L)^2 T\!L_n(y;L)-1$, which individually satisfy the boundary conditions, are related to  the rational Chebyshev functions $T\!L_n(y;L)$ on the semi-infinite interval $0<y<\infty$ via the ``basis recombination" \cite{boyd2001chebyshev}. 
The adjustable parameters $L_x$ and $L_z$ are set to $1$ for solving $\tilde{G}(x,z)$
while we use $L_x = 16$ and $L_z = 1$ for $\tilde{H}(x,z)$ to better stabilize the $x$-dependence of $H$. The number of basis functions  $N$ is varied in the range $60\le N\le 120$, and the variation in the results is used as an estimate of systematic errors for each value of $k$. 
Libraries \texttt{SciPy}~\cite{Virtanen:2019joe}, \texttt{Eigen}~\cite{eigen} and
\texttt{Boost.Multiprecision}~\cite{boost} have been  helpful for these numerical analyses. 

The solutions just described depend on the input value of $\rho$. For the instanton solution, $\rho$ is given by (\ref{parameter2}). Since we go beyond the instanton approximation, $\rho$ needs to be recalculated accordingly. For our new solution, the nucleon mass 
(\ref{saka}) takes the form   
\beq
M(\rho)
&=&  2\kappa \int_0^\infty  \frac{ q^2dq}{2\pi^2} \int_{0}^\infty dz  \Biggl[\frac{h(z)}{2}\left(q^2 \widehat{A}^2_0(q,z) + (\partial_z \widehat{A}_0(q,z))^2\right) \Biggr]\nn &&+ 8\pi \kappa  (4\pi^2\rho^2)^2\int_0^\infty  x^2 dx \int_0^\infty dz  \nn 
&& \quad \times\Biggl[h(z)\left({\cal E}^2+({\cal D}+{\cal F})^2+\frac{1}{2}({\cal D}+{\cal I})^2\right) + k(z)({\cal A}^2+{\cal C}^2+({\cal A}+{\cal B})^2)\Biggr].
\eeq
We find $M(\rho)$ has a minimum at  (compare to (\ref{parameter2}))
\beq
\rho^* \approx 2.26 , \qquad  M(\rho^*) \approx 1.31M_{KK}.
\eeq
  We have thus evaluated the integrals (\ref{p1}) and (\ref{dsu2}) using the solutions $\widehat{A}_0$, $G$ and $H$  with $\rho =\rho^*$, and extrapolated them to $k\to 0$ to obtain 
\beq
D^{\rm U(1)}(0) \approx 0.543, \qquad D^{\rm SU(2)}(0)\approx -0.685\pm 0.022,   \label{su2cont}
\eeq  
where the errors for the U(1) part are negligibly small. 
 After a cancellation between the positive U(1)  and negative SU(2) contributions, the total D-term 
 \beq
 D(0)= D^{\rm SU(2)}(0)+ D^{\rm U(1)}(0)\approx -0.140 \pm 0.022,
 \eeq
 turns out to be slightly negative. 
That the U(1) contribution is positive is  intuitively  easy to understand. The U(1) field is analogous to the static electric field $\vec{E}\sim \vec{r}/r^3$ of a point charge. The energy momentum tensor (Maxwell's stress tensor) of a point charge takes the form
\beq
T_{ij} \sim -E_iE_j+\frac{\delta_{ij}}{2}\vec{E}^2 = -\frac{1}{r^4}\left(\frac{r^ir^j}{r^2}-\frac{\delta_{ij}}{3}\right) +\frac{\delta_{ij}}{6r^4}. 
\eeq
The `pressure'  $p(r)\sim 1/(6r^4)$ (see (\ref{shear})) is everywhere positive and hence the D-term $D\sim \int d^3r r^2 p(r)$ is also positive (actually divergent). On the other hand, the SU(2) fields may be thought of as the `pion cloud' in traditional hadron physics. In the chiral quark soliton model, it has been argued that the pion cloud is  responsible for making the  D-term negative  \cite{Goeke:2007fp}. Our result is  consistent with this argument, although in the present approach the `cloud' is made up of  not only  pions but also infinitely many vector and axial-vector mesons. Incidentally, if we neglect the non-Abelian commutator terms (terms proportional to $\rho^2$ in (\ref{abc})), the SU(2) contribution also becomes positive. 

Our result provides a new perspective on the stability of nucleons in holographic QCD.  If we interpret the D-term as a measure of  outward radial force, the U(1) and SU(2) fields generate  positive and negative forces, respectively, and they tend to expand and shrink the system. This is entirely analogous to the calculation of the nucleons mass as briefly mentioned below (\ref{stable}) and discussed in detail in \cite{Hata:2007mb}. Namely, the U(1) field (iso-singlet mesons, in particular the $\omega$ meson) tends to expand the nucleon by preferring  large instanton sizes $\rho\to \infty$, and this is counterbalanced by the SU(2) fields (iso-vector mesons  $\pi,\rho,a_1,\cdots$) which prefer small sizes $\rho\to 0$. Neither one of them alone can stabilize the nucleon. Thus, there seems to be a direct link between the stability arguments in terms of the nucleon's mass and D-term when they are decomposed into contributions from different subsystems.  On the other hand, the present discussion does not indicate whether the sign of the total $D=D^{\rm U(1)}+D^{\rm SU(2)}$, which happens to be slightly negative, is of particular significance regarding stability (cf.,  \cite{Metz:2021lqv,Ji:2021mfb}).

\section{Coupling to gravity}

In gauge/string duality, the proper method to calculate the field theory expectation value of the energy momentum tensor has been established 
\cite{Balasubramanian:1999re,deHaro:2000vlm}. 
In this section, we apply the framework of  holographic renormalization developed in \cite{deHaro:2000vlm,Kanitscheider:2008kd} to the calculation of the form factor $D(|\vec{k}|)$ and elucidate its connection to the glueball spectrum. We shall also explain why the classical calculation in the previous section is valid only at $|\vec{k}|=0$.   For previous attempts in bottom-up holographic models, see \cite{Mamo:2019mka,Mamo:2021krl}.

\subsection{Setup}

The basic idea of our holographic calculation is that  matter fields in the `bulk' perturb the metric, and this `wake' is propagated to the boundary and recorded as the field theory expectation value $\langle T_{\mu\nu}\rangle$. In bottom-up  holographic QCD models, gravity is confined in  the same five-dimensional (deformed) anti-de Sitter spaces where the matter fields live. However, the situation is different in our top-down model. 
The action (\ref{action}) is a low-energy effective theory on the `flavor' D8 branes embedded in a ten-dimensional curved space-time in type IIA supergravity. The latter is further derived from   eleven-dimensional supergravity (M-theory) on doubly Wick rotated AdS$_7$ black hole $\times S^4$  \cite{Witten:1998xy}
\beq
ds^2 = \frac{r^2}{L^2}[f(r) d\tau^2 -dx_0^2+ dx_1^2 + dx_2^2+dx_3^2+dx_{11}^2] + \frac{L^2}{r^2}\frac{dr^2}{f(r)} + \frac{L^2}{4}d\Omega_4^2,  \label{bh}
\label{fe}
\eeq
after compactifying the eleventh dimension $x_{11}$ on a circle of radius 
\beq
R_{11}=g_sl_s = \frac{\lambda}{2\pi N_c M_{KK}}. \label{compact11}
\eeq
In (\ref{bh}) we defined   
\beq
 f(r)=1-\frac{R^6}{r^6}, \qquad R=\frac{L^2M_{KK}}{3}, \qquad L^3=8\pi g_sN_cl_s^3, 
\eeq
where $g_s$ is the string coupling and $l_s=\sqrt{\alpha'}$ is the string length. 
The following relation is useful
\beq
\lambda N_c=\frac{L^6M_{KK}}{32\pi g_s l_s^5}. \label{useful}
\eeq
The radial coordinate $r$ ($\infty > r>R$) is related to $z$ in (\ref{action}) as 
\beq
z=\pm \sqrt{\frac{r^6}{R^6}-1}. \label{z}
\eeq
 Below we shall use $r$ and $z$ interchangeably keeping in mind that they are related as (\ref{z}).

The AdS$_7$ black hole geometry is sourced by the $N_c$ D4 branes spanning the coordinates $(x^{\mu=0,1,2,3},\tau)$. In this background, $N_f$ D8 branes are placed at $\tau=0$ and $\tau=\pi/M_{KK}$ corresponding to the $z>0$ and $z<0$ branches, respectively, which are smoothly connected at $r=R$. We adopt the probe approximation $N_c\gg N_f$ and the backreaction of the geometry due to the D8 branes are neglected. The $S^4$ part of the D8 branes has been already integrated out in (\ref{action}). 
The dual field theory on the boundary of AdS$_7$ at $r\to \infty$ is a six-dimensional conformal field theory coupled with four-dimensional fermions at the location of the D8/$\overline{\rm D8}$ branes.\footnote{This six-dimensional conformal field theory is the ${\cal N}=(2,0)$ superconformal field theory realized on $N_c$ M5 branes extended along $x^{0,1,2,3,11}$ and $\tau$ directions, which are the M-theory lift of the D4 branes in type IIA string theory. On the other hand, the M-theory lift of a D8 brane is not well-understood. We just use the M-theory description as a convenient notation to organize the fields in type IIA supergravity.} After compactifying the $x_{11}$ direction (\ref{compact11}) and furthermore the $\tau$-direction on a circle  $\tau\sim \tau+2\pi/M_{KK}$ with supersymmetry breaking boundary conditions, the theory becomes a four-dimensional, confining $SU(N_c)$ Yang-Mills theory coupled with $N_f$ Dirac fermions, that is $SU(N_c)$ QCD with $N_f$ flavors, at low energies. Our task is to calculate the induced energy momentum tensor $\langle T_{ij}\rangle$ in this boundary theory sourced by the soliton, which corresponds to the nucleon, living on the D8 branes.

While the method of holographic renormalization has been extended to non-conformal theories    \cite{Kanitscheider:2008kd}, for our purposes, it is more convenient to work in the eleven-dimensional (or seven-dimensional, after reducing on $S^4$) setting (\ref{bh}) instead of ten-dimensional type-IIA supergravity with the dilaton. In this setting, we calculate the metric fluctuation caused by the bulk soliton ($x^M=(x^{0,1,2,3},\tau,r,x_{11}$)) 
\beq
\delta g_{MN}(\tau,x_{11},x,r) &=& \kappa^2_{7}  \int d\tau'dx'_{11}d^4x'dr'\sqrt{-G'_{(7)}} \nn 
&& \qquad \times G_{MNAB}(\tau-\tau',x_{11}-x'_{11},x-x',r,r')   {\cal T}^{AB}(\tau',x'_{11},x',r'), \label{flu}
\eeq
in the axial gauge
\beq
\delta g_{M\scalebox{0.9}{$r$}}=0, \label{axi}
\eeq
and study the behavior of the four-dimensional components $\delta g_{\mu\nu}$ ($\mu,\nu=0,1,2,3$) near the boundary $r\to \infty$. The expectation value of the energy momentum tensor is then proportional to the coefficient of the $1/r^4$ term integrated over the extra dimensions
\beq
 \int_{x_{11},\tau} \delta g_{\mu\nu} \sim \frac{C_{\mu\nu}}{r^4}+\cdots ,\qquad \langle T_{\mu\nu}\rangle \propto C_{\mu\nu}.  \label{holoreno}
\eeq
(See \cite{deHaro:2000vlm} for the precise prescription.) 
In (\ref{flu}),  $G_{MNAB}$ is the graviton propagator and $G'_{(7)}=-(r'/L)^{10}$ is the determinant of the AdS part of the metric.  
  ${\cal T}^{AB}$  ($A,B=0,1,2,3,\tau,r,11$) is  the  seven-dimensional energy momentum tensor (already integrated over $S^4$) of the soliton. $\kappa_7^2$ is related to the eleven-dimensional gravitational constant as 
  \beq
  \kappa_{11}^2 = {\rm Vol}(S^4) \kappa_{7}^2 = \frac{8\pi^2}{3}\left(\frac{L}{2}\right)^4 \kappa_7^2.
  \eeq
For a static soliton, ${\cal T}^{AB}$ does not depend on $x'_0$ nor $x'_{11}$. Going to the Fourier space, we get  
\beq
\delta g_{MN}(\vec{k},r)&\equiv&
\int d\tau dx_{11} d^3\vec{x}\,e^{i\vec{k}\cdot \vec{x}}\delta g_{MN}(\tau,x_{11},\vec{x},r)\nonumber\\
&=& \kappa^2_{7}\int dr' \sqrt{-G'_{(7)}}G_{MNAB}(\vec{k},r,r'){\cal T}^{AB}(\vec{k},r'), \label{space}
\eeq
where $G(\vec{k},r,r')$ is the momentum space graviton propagator  with $k_{\scalebox{0.9}{$\tau$}}=k_{11}=k_0=0$, and we defined 
\beq
{\cal T}^{AB}(\vec{k},r')\equiv 2\pi R_{11}\int d\tau' {\cal T}^{AB}(\tau',\vec{k},r').
\label{intT}
\eeq
 The $\tau'$-integral is trivial because the configuration of D8 branes described above instructs us to write, in the $z$-coordinates, 
\beq
{\cal  T}^{AB}(\tau',\vec{k},z)=\frac{ {\cal T}^{AB}(\vec{k},|z|) }{4\pi R_{11}}\left(\delta(\tau')+\delta(\tau'-\pi/M_{KK})\right). \label{eleven}
 \eeq

 The bulk energy momentum tensor
 ${\cal T}_{AB}$ can be obtained by varying the D8 brane action with respect to the eleven dimensional metric (\ref{bh}) 
 \beq
{\cal T}_{AB}=-\frac{2}{\sqrt{-G_{(11)}}}\frac{\delta S_{D8}}{\delta G_{(11)}^{AB}}{\rm Vol}(S^4),
\eeq
where we only keep the quadratic terms in the field strength tensor in the D8 brane action 
\beq
S_{D8} \approx  - C\int d^4x dr d^4\Omega_4 e^{-\Phi}\sqrt{-\tilde{g}}\frac{1}{4}\tilde{g}^{mn}\tilde{g}^{ab}{\rm tr}[F_{ma}F_{nb}], \label{d8b}
\eeq
with $C=(64\pi^6l_s^5)^{-1}$. $\tilde{g}$ is the induced metric on the D8-brane 
\beq
\tilde{g}_{ab} dx^a dx^b=\frac{
r^3}{L^3}[-dx_0^2 + (d\vec{x})^2] + \frac{r}{L}\left(\frac{L^2}{r^2f(r)} + f(r) \left(\frac{\partial \tau}{\partial r}\right)^2\right)
dr^2 + \frac{r}{L}\frac{L^2}{4}d\Omega_4^2,
\eeq
with a nontrivial dilaton field $e^{-\Phi}=\frac{1}{g_s}(L/r)^{3/2}$. (In the present case, $\partial \tau /\partial r=0$.) 
In the eleven-dimensional notation, (\ref{d8b}) takes the form 
\beq
S_{D8}&=&-\frac{C}{2\pi R_{11}g_s} \int d\tau dx_{11} d^4xdrd^4\Omega_4\left(\delta(\tau)+\delta(\tau-\pi/M_{KK})\right) \nn && \qquad \qquad \qquad \times \frac{\sqrt{-G_{(11)}}}{\sqrt{G_{\scalebox{0.9}{$\tau\tau$}}}} \frac{1}{4}G^{AB}G^{MN}{\rm tr}\left[F_{AM}F_{BN}\right]+\cdots. \label{118}
\eeq
This leads to 
 \beq
2\pi R_{11}\int d\tau' \sqrt{-G'_{(7)}}{\cal T}_{\mu\nu}
&=&\frac{\pi^2Cr'^2L^2}{3g_s\sqrt{f(r')}}
{\rm tr}\Biggl[F_{\mu\rho}F_{\nu}^{\ \rho}+(1+z^2)^{\frac{4}{3}}F_{\mu z}F_{\nu z} \nn 
&& \qquad \qquad-\frac{\eta_{\mu\nu}}{4}\left(F_{\rho\sigma}F^{\rho\sigma} + 2(1+z^2)^{\frac{4}{3}}F_{\rho z}F^\rho_{\ z}\right)\Biggr] ,\label{use} \\  
2\pi R_{11} \int d\tau' \sqrt{-G'_{(7)}}{\cal T}_{11,11}&=& \frac{\pi^2 C r'^2L^2}{3g_s \sqrt{f(r')}}{\rm tr}\left[-\frac{1}{4}F_{\rho\sigma}F^{\rho\sigma} -\frac{(1+z^2)^{\frac{4}{3}}}{2}  F_{\rho z}F^{\rho}_{\  z}\right],\label{1111} \\ 
2\pi R_{11} \int d\tau' \sqrt{-G'_{(7)}}{\cal T}_{\mu z }&=& \frac{\pi^2 C r'^2L^2}{3g_s \sqrt{f(r')}}{\rm tr}\left[F_{\mu\rho}F_z^{\ \rho}\right], \label{zz}
\\
2\pi R_{11} \int d\tau' \sqrt{-G'_{(7)}}{\cal T}_{zz}&=& \frac{\pi^2 C r'^2L^2}{3g_s \sqrt{f(r')}}{\rm tr}\Biggl[\frac{1}{2}F_{z\rho}F_z^{\ \rho} -\frac{1}{4(1+z^2)^{\frac{4}{3}}} F_{\rho\sigma}F^{\rho\sigma}\Biggr],
\label{tzz}
\eeq
with 
\beq
{\cal T}_{\mu \scalebox{0.9}{$r$}} = \frac{\partial z}{\partial r} {\cal T}_{\mu z} =\frac{3r^5}{zR^6}{\cal T}_{\mu z} ,\qquad {\cal T}_{\scalebox{0.9}{$rr$}} = \frac{9r^{10}}{z^2R^{12}}{\cal T}_{zz}. \label{convert}
\eeq
[For simplicity only the SU(2) part is shown.]  
 It is important to notice that (\ref{118}) is independent of $G_{\scalebox{0.9}{$\tau\tau$}}$, hence  ${\cal T}_{\scalebox{0.9}{$\tau\tau$}}=0$ for the soliton configuration. This is simply   because the D8/$\overline{\rm D8}$ branes do not extend in the $\tau$-direction. On the other hand, ${\cal T}_{11,11}$ is nonvanishing even though the D8/$\overline{\rm D8}$ branes do not extend in the $x_{11}$ direction. This can be easily understood as the coupling between the gauge fields and the dilaton field in the type IIA description in ten dimensions.  

\subsection{Glueballs in the AdS$_7$ black hole}

In this paper, we do not attempt to compute (\ref{space}) in its full glory because we do not have the exact (numerical) solution of the soliton configuration. Instead, assuming that the latter is known,  we will analyze the bulk Einstein equation in detail and establish the connection between the gravitational form factors and the known glueball spectrum of this theory \cite{Constable:1999gb,Brower:2000rp} (see, also,  \cite{Hashimoto:1998if,Hashimoto:2007ze,Brunner:2015oqa}).  

 The relevance of glueballs to the present problem can be understood  as follows.  Since  the spatial part of the field theory  energy momentum tensor is transverse $k^iT_{ij}=0$, near the boundary the metric fluctuation  must also be transverse 
$k^i\delta g_{ij}=0$ and can be parameterized as \footnote{Near the boundary $r\to \infty$, the metric fluctuation in the $d=6$ dimensional subspace $x^a=(x^{0,1,2,3,11},\tau)$ can be generically written as 
\beq
\delta g_{ab}\approx  \delta g_{ab}^{\rm TT} + \frac{k_a k_c \delta g_{b}^{c} + k_b k_c \delta g_{a}^c  }{k^2}- \frac{k_a k_b k_c k_d \delta g^{cd} }{(k^2)^2} +\frac{1}{d-1}\left(\eta_{ab}-\frac{k_a k_b}{k^2}\right)\left(\delta g^c_c-\frac{k_c k_d \delta g^{cd}}{k^2}\right),
\eeq
in momentum space. Taking   $k^a=\delta^a_i k^i$ and imposing the condition $k_i\delta g^i_j=0$ we arrive at (\ref{traceless}).  
}
\beq
 \delta g_{ij} \approx \delta g_{ij}^{\rm TT} +\frac{1}{5} \left(\delta_{ij}-\frac{k_ik_j}{\vec{k}^2} \right) \delta g^a_a. \qquad (r\to \infty)\label{traceless}
 \eeq
 $\delta g^{\rm TT}$ is the so-called transverse-traceless (TT) part where it is understood that `trace'  is taken in six-dimensions $x^a=(x^{\mu=0,1,2,3},x^{11},\tau)$. As for the trace part, since we neglect the backreaction of the D8/$\overline{\rm D8}$ branes in the probe approximation,\footnote{The backreacted geometry for the present problem has been calculated in \cite{Burrington:2007qd}.}  the geometry is asymptotically AdS with no dilaton. In the  axial gauge, the Einstein equation then automatically requires that  $\delta g^a_a\approx 0$ in the asymptotically AdS regime.\footnote{If one works in the type IIA setup or in AdS$_5$/QCD$_4$ models, one must include this term and solve the Einstein equation coupled to the dilaton field \cite{Kanitscheider:2008kd,Mamo:2019mka}. } Therefore, the trace term can be neglected and the D-term entirely originates from the TT modes  of the AdS$_7$ black hole. 
Glueballs are nothing but the normalizable TT modes of the  linearized Einstein equation. 
They can be classified according to spin under the rotation group SO(3) in physical space and interpreted as the actual spectrum of glueballs on the boundary field theory.
Among the 14 independent TT modes, those relevant to the present discussion are referred to as `T$_4$' and `S$_4$' in the classification of \cite{Brower:2000rp}.\footnote{In this paper, we do not consider metric fluctuations on $S^4$. Actually, if one does that, there exists another $0^{++}$  glueball mode called `L$_4$'  \cite{Hashimoto:1998if,Brower:2000rp} which can be sourced by the  $S^4$ part of the energy momentum tensor (obtained similarly to ${\cal T}_{11,11}$ in (\ref{1111})). However, this is not a TT mode, and one can show that it does not contribute to the boundary energy momentum tensor.}   
  The T$_4$ mode consists of  $2^{++}$, $1^{++}$ and $0^{++}$ glueballs. They are transverse and traceless  in five dimensions $x^{0,1,2,3,11}$ and have vanishing components in the other dimensions.  Despite the difference in spin, all these glueballs obey the same bulk equation of motion,  namely the massless Klein-Gordon equation on the AdS$_7$ black hole. Therefore, their masses are degenerate. The T$_4 (2^{++})$ glueballs are transverse-traceless already in four dimensions $x^{0,1,2,3}$. Among the five independent 2$^{++}$ polarizations, the following mode stands out as it features the same spatial tensor structure as in (\ref{br})
  \beq
  \delta g^{{\rm T}_4}_{\mu\nu}(2^{++}) \sim \begin{pmatrix} 2 & \ell_j \\ \ell_i & \delta_{ij} -\frac{k_ik_j}{\vec{k}^2}  \end{pmatrix},
  \label{t4}
  \eeq 
  where $\ell_i$ satisfies $k^i\ell_i=0$.
  The metric fluctuation corresponding to the T$_4 (0^{++})$ glueballs are traceless in five dimensions     \cite{Constable:1999gb,Brower:2000rp}
 \beq
 \delta g^{{\rm T}_4 }_{ab}(0^{++}) \sim \begin{pmatrix} \delta g_{\mu\nu} & \delta g_{\mu, 11} \\ \delta g_{11,\mu} & \delta g_{11,11} \end{pmatrix} \sim  \begin{pmatrix} \eta_{\mu\nu} - \frac{k_\mu k_\nu}{k^2} & 0 \\ 0 & -3 \end{pmatrix},
 \label{0++}
 \eeq
 with $k^\mu=(0,\vec{k})$. The spatial part again contains the structure $k_ik_j-\delta_{ij}\vec{k}^2$ we look for. Note that the largest component is in the eleventh dimension. Therefore, the T$_4(0^{++})$ mode may be thought of as the counterpart of the dilaton in type IIA supergravity.

On the other hand, the S$_4$ mode consists only of $0^{++}$ glueballs. Their masses are not degenerate with the T$_4$ glueballs, and actually the lightest scalar glueball belongs to this class. In \cite{Constable:1999gb} this solution was dubbed  `exotic' because it has a nonzero  component in the $\tau$-direction
\beq
\delta g_{ab}^{{\rm S}_4} \sim \begin{pmatrix} \delta g_{\mu\nu} &  & \\   & \delta g_{11,11} &  \\ 
 & & \delta g_{\scalebox{0.9}{$\tau\tau$}}\end{pmatrix} \sim  \begin{pmatrix} \eta_{\mu\nu} - \frac{k_\mu k_\nu}{k^2}   && \\  & 1 &  \\ 
 & & -4 \end{pmatrix}. 
 \label{simple}
 \eeq 
 Far away from the boundary, the metric fluctuation cannot be written in this simple form, and it is no longer possible to literally maintain the  `transverse-traceless' condition. Besides, the explicit solution constructed in  \cite{Constable:1999gb} features nonvanishing components in the $r$-direction (see (\ref{s4met}) below).

We thus see that there are three different types of glueballs that can potentially contribute to the D-term. 
Accordingly, the bulk energy momentum tensor (\ref{use})-(\ref{1111}) can be decomposed  as\footnote{In the following, the trivial $\tau$-integral (cf. (\ref{space}), (\ref{intT})) is understood both for the bulk energy momentum tensor ${\cal T}^{AB}$ and the metric fluctuation $\delta g_{MN}$.} 
\beq
{\cal T}_{AB} = {\cal T}^{{\rm T}(2)}_{AB} + {\cal T}_{AB}^{{\rm T}(0)} +{\cal T}_{AB}^{\rm S}+{\cal T}_{AB}^{\rm other}, \label{other}
\eeq
where ${\cal T}^{{\rm T}(2)}_{AB}$ is the part that sources the  T$_4 (2^{++})$ excitation (\ref{t4}), etc. ${\cal T}_{AB}^{\rm other}$ is the remainder which does not contribute to  $\langle T_{\mu\nu}\rangle$ on the boundary. (See (\ref{Tother})--(\ref{tautau}).) Since each TT mode satisfies a decoupled  equation, near the boundary we can write, schematically, 
\beq
 \delta g_{\mu\nu} \sim G^{{\rm T}(2)}_{\mu\nu AB}{\cal T}^{AB}_{{\rm T}(2)} + G^{{\rm T}(0)}_{\mu\nu AB}{\cal T}^{AB}_{{\rm T}(0)} +G^{{\rm S}}_{\mu\nu AB}{\cal T}^{AB}_{{\rm S}}. \label{s4t4}
\eeq 
The boundary energy momentum tensor $\langle T_{ab}\rangle$ of the six-dimensional field theory is therefore given by a linear combination
\beq
\langle T_{ab}\rangle &=& t_2(\vec{k}^2) \begin{pmatrix} \begin{pmatrix} 2\vec{k}^2  &  \vec{k}^2\ell_j\\
\vec{k}^2\ell_i &  \vec{k}^2\delta_{ij}-k_ik_j \end{pmatrix} & & \\ & & 0 & \\ & & & 0   \end{pmatrix} + t_0(\vec{k}^2) \begin{pmatrix} k^2\eta_{\mu\nu}-k_\mu k_\nu &   & \\  & -3k^2 & \\ & & 0 \end{pmatrix}  \nn && + s_0(\vec{k}^2)   \begin{pmatrix} k^2\eta_{\mu\nu}-k_\mu k_\nu &   & \\  & k^2 & \\ & & -4k^2 \end{pmatrix} \nn
&=& \begin{pmatrix} (2t_2-t_0-s_0)\vec{k}^2 & \vec{k}^2\ell_j& &  \\ 
\vec{k}^2\ell_i & (t_2+t_0+s_0)(\vec{k}^2\delta_{ij}-k_ik_j) & & \\ 
  &  & (-3t_0+s_0)\vec{k}^2 & \\ & & & -4s_0\vec{k}^2 \end{pmatrix}. \label{boundary}
\eeq
Comparing the four-dimensional part of this  to (\ref{br}) or (\ref{can}), we can identify\footnote{For the sake of generality, here we temporarily restore the spin-1/2 nature of the nucleon. In the present treatment using the classical bulk energy momentum tensor (\ref{use}) with spherically symmetric gauge configurations, the dependence of the energy momentum tensor on the spin states of the nucleon is not captured and $\ell_i$ vanishes.} 
\beq
t_2 &=& \frac{M}{3}\left(\frac{A}{\vec{k}^2}-\frac{B}{4M^2}\right)\delta_{s's},\nn
t_2\ell_i\vec{k}^2&=&\frac{A+B}{4}\epsilon_{ijk}(ik^j)\sigma^k_{s's}
\nn
t_0+s_0 &=& -\frac{M}{3}\left(\frac{A}{\vec{k}^2}-\frac{B}{4M^2}+\frac{3D}{4M^2}\right)\delta_{s's}.
\label{kpole}
\eeq
Note that $t_2$ and $t_0+s_0$ both  have a pole $1/\vec{k}^2$ but they cancel in the sum. 
The components of (\ref{boundary}) are then given by 
\beq
\langle T_{00}\rangle &=& (2t_2-t_0-s_0)\vec{k}^2 = M\left(A-\frac{B\vec{k}^2}{4M^2}+\frac{D\vec{k}^2}{4M^2}\right)\delta_{s's},\\
\langle T_{0i}\rangle&=&
t_2\ell_i\vec{k}^2
=\frac{A+B}{4}\epsilon_{ijk}(ik^j)\sigma^k_{s's},\\
\langle T_{ij}\rangle &=& (t_2+t_0+s_0)(\vec{k}^2\delta_{ij}-k_ik_j) = \frac{D}{4M} (k_ik_j-\vec{k}^2\delta_{ij})\delta_{s's}, \label{cancel}
\eeq
in agreement with (\ref{br}). 
Moreover, 
the traceless condition in six dimensions leads to the QCD trace anomaly relation in four dimensions 
\beq
-\langle T^\mu_\mu\rangle = \langle T^{11}_{11}\rangle +\langle T_{\scalebox{0.9}{$\tau$}}^{\scalebox{0.9}{$\tau$}}\rangle= M\left(A-\frac{B\vec{k}^2}{4M^2}+3\frac{D\vec{k}^2}{4M^2}\right)\delta_{s's}.  \label{anomaly}
\eeq

In order to determine the parameters $t_0,t_2,s_0$, one needs to fully solve the linearized Einstein equation with the (numerically obtained) bulk energy momentum tensor. This will be discussed in the next subsection. A naive expectation, however,  is that the S$_4$ mode decouples or is  suppressed compared to the T$_4$ modes, namely, $|s_0|\ll |t_{0,2}|$. Indeed,  there have been arguments  in the literature \cite{Constable:1999gb,Brunner:2015oqa}  that the S$_4$ glueball states may not survive in the `continuum   limit' $M_{KK}\to \infty$ keeping meson masses fixed and including  the stringy $1/\lambda$ corrections to all orders.  The corresponding metric fluctuation (\ref{simple}) has the largest  component in the $\tau$-direction whose  compactification radius $1/M_{KK}$ shrinks to zero in this limit. Such arguments are also practically motivated since  
 there is an excess of $0^{++}$ glueballs from holography compared with  lattice QCD results 
 \cite{Brower:2000rp,Brunner:2015oqa}.  We however note  that  there is no symmetry argument which prevents the S$_4$ mode from surviving the continuum limit. 
 In this paper, we will not try to include the stringy corrections, but work in the supergravity approximation without taking the continuum limit. Within this approximation, we will see that the S$_4$ mode actually contributes to the gravitational form factors.

\subsection{Solving the Einstein equation}

Let us investigate in detail whether and how the various glueball states described in the previous subsection couple to the soliton.   For this purpose, we turn to the linearized Einstein equation  
 \beq
{\cal H}_{AB}&\equiv& \nabla^2 \delta g_{AB}+ \nabla_A\nabla_B \delta g^{C}_C -\nabla^C(\nabla_A \delta g_{BC} + \nabla_B \delta g_{AC})-\frac{12}{L^2}\delta g_{AB}\nn
&=& -2\kappa_7^2 \left({\cal T}_{AB}-\frac{g_{AB}}{5}{\cal T}^C_C\right), \label{einstein4}
\eeq
where the covariant derivative and raising/lowering of indices are calculated with respect to the background metric $g_{AB}$ in (\ref{bh}). 
We do not impose $\delta g^C_C\approx 0$ at this point, since this condition holds only near the boundary $r\to \infty$. After eliminating ${\cal T}^C_C$ by taking a trace,  (\ref{einstein4}) can be cast into an equivalent form  
\beq
\overline{\cal H}_{AB}={\cal H}_{AB}-\frac{g_{AB}}{2}{\cal H}^C_C = -2\kappa_7^2 {\cal T}_{AB}. \label{equivalent}
\eeq
Let us first substitute the following parametrization relevant to  the T$_4 (2^{++})$ mode  (see (\ref{t4})),
\beq
&& \delta g^{{\rm T}(2)}_{00} =2\frac{r^2}{L^2} h^{{\rm T}(2)}(r,k), \nn 
 &&\delta g^{{\rm T}(2)}_{ij}= \frac{r^2}{L^2}  \left(\delta_{ij}-\frac{k_i k_j}{\vec{k}^2}\right)h^{{\rm T}(2)}, 
 \label{00s}
 \eeq
  we find, for $k^\mu=(0,\vec{k})$, 
 \beq
 && \overline{\cal H}_{00}=2\frac{r^2}{L^2}\nabla^2 h^{{\rm T}(2)}, \nn 
 &&\overline{\cal H}_{ij}= \frac{r^2}{L^2} \left(\delta_{ij}-\frac{k_i k_j}{\vec{k}^2}\right)\nabla^2 h^{{\rm T}(2)}, \label{t2h}
 \eeq
 with all the other components vanishing.  
Here, $\nabla^2$ is the massless Klein-Gordon operator
 \beq
 \nabla^2=  \frac{1}{L^2r^5}\partial_r\bigl( (r^7-rR^6)\partial_r\bigr) -\frac{L^2k^2}{r^2},
 \eeq 
 and the propagation is diagonal in Lorentz indices.  Naturally, this mode can be sourced by a part of the bulk ${\cal T}_{\mu\nu}$ that has the same tensorial structure as in (\ref{t2h}).  
  As for the T$_4 (0^{++})$ mode, we use (see (\ref{0++}))
 \beq
 &&\delta g^{{\rm T}(0)}_{\mu\nu}= \frac{r^2}{L^2}  \left(\eta_{\mu\nu}-\frac{k_\mu k_\nu}{k^2}\right)h^{{\rm T}(0)}(r,k), \nn
 && \delta g^{{\rm T}(0)}_{11,11} =-3\frac{r^2}{L^2}h^{{\rm T}(0)}, \label{0+g}
 \eeq
 and find that 
 \beq
 &&\overline{\cal H}_{\mu\nu}= \frac{r^2}{L^2} \left(\eta_{\mu\nu}-\frac{k_\mu k_\nu}{k^2}\right)\nabla^2 h^{{\rm T}(0)}, \nn
 && \overline{\cal H}_{11,11}= -3\frac{r^2}{L^2}\nabla^2 h^{{\rm T}(0)}. \label{t0h}
 \eeq
Again there is no mixing of Lorentz indices. This explains the above-mentioned degeneracy between the T$_4(2^{++})$ and T$_4(0^{++})$ glueballs. Turning to ${\cal T}_{AB}$ on the right hand side, we see from (\ref{1111}) that the component ${\cal T}_{11,11}$ is nonvanishing. Therefore, the soliton can act as a source for  the T$_4(0^{++})$ mode (or the dilaton in the type IIA language).

Next,  we substitute the following S$_4(0^{++})$ metric fluctuation     \cite{Constable:1999gb}\footnote{Note that (\ref{s4met}) is not expressed in the axial gauge (\ref{axi}). However, this does not affect the discussion below because  $\overline{\cal H}_{AB}$ is gauge invariant and $\delta g_{M\scalebox{0.9}{$r$}}$ rapidly vanishes in the large $r$ limit.}   into the left hand side of (\ref{equivalent})
\begin{equation}
\begin{split}
&\delta g^{{\rm S}}_{\scalebox{0.9}{$\tau\tau$}}=-f(r)\frac{r^2}{L^2}h^{\rm S}(r,k), \\
&\delta g^{{\rm S}}_{\mu\nu}= \frac{r^2}{4L^2}\left[\eta_{\mu\nu}- \frac{k_\mu k_\nu}{k^2} - \frac{12R^6}{5r^6-2R^6}\frac{k_\mu k_\nu}{k^2}\right]h^{\rm S}, \\
&\delta g^{{\rm S}}_{11,11} = \frac{r^2}{4L^2}h^{\rm S}, \label{s4met}\\
& \delta g^{{\rm S}}_{\scalebox{0.9}{$rr$}} =-\frac{L^2}{r^2f(r)}\frac{3R^6}{5r^6-2R^6}h^{\rm S}, \\ 
& \delta g^{{\rm S}}_{\scalebox{0.9}{$r$}\mu}= \frac{90r^7 R^6}{L^2(5r^6-2R^6)^2}\frac{ik_\mu}{k^2}h^{\rm S}.
\end{split}
\end{equation}
 The result is 
  \begin{equation}
 \begin{split}
 &\overline{\cal H}_{\scalebox{0.9}{$\tau\tau$}}= -\frac{r^2f(r)}{L^2}\frac{5(r^6-R^6)}{5r^6-2R^6}{\cal H}^{\rm S},\\
 & \overline{\cal H}_{\mu\nu} =\frac{r^2}{L^2} \frac{5(r^6+2R^6)}{4(5r^6-2R^6)}\left(\eta_{\mu\nu}-\frac{k_\mu k_\nu}{k^2}\right) {\cal H}^{\rm S} , \\ 
 & \overline{\cal H}_{11,11} = \frac{r^2}{L^2}\frac{5(r^6+2R^6)}{4(5r^6-2R^6)}  {\cal H}^{\rm S}, \\ 
 &\overline{\cal H}_{\scalebox{0.9}{$rr$}} =0, \\ 
 &\overline{\cal H}_{\scalebox{0.9}{$r$}\mu}=0, \label{s4metH} \end{split}
 \end{equation}
 where
 \beq
 {\cal H}^{\rm S}&\equiv& \left(  \frac{1}{L^2r^5}\partial_r\bigl( (r^7-rR^6)\partial_r\bigr) -\frac{L^2k^2}{r^2}+\frac{432R^{12}}{L^2(5r^6-2R^6)^2}\right)h^{\rm S} \nn 
 &=& \left( \nabla^2  +\frac{432R^{12}}{L^2(5r^6-2R^6)^2}\right) h^{\rm S}.
 \label{extra}
 \eeq
The differential operator in (\ref{extra})  agrees with the one derived in  \cite{Constable:1999gb,Brower:2000rp} for the S$_4$ mode. The following identity will be very useful later
 \beq
&&\frac{5(r^6+2R^6)}{5r^6-2R^6} \left(\frac{1}{r^3}\partial_r (r(r^6-R^6)\partial_r) + \frac{432r^2R^{12}}{(5r^6-2R^6)^2}\right) \nn 
&& = \frac{1}{r^3} \partial_r \left(r(r^6-R^6)\frac{5(r^6+2R^6)}{5r^6-2R^6}\left(\partial_r+ \frac{72r^5R^6}{(5r^6-2R^6)(r^6+2R^6)}\right)\right) \nn 
&& = \frac{1}{r^3}\partial_r\left(r(r^6-R^6)\left(\partial_r + \frac{144r^5R^6}{(5r^6-2R^6)(r^6+2R^6)}\right) \frac{5(r^6+2R^6)}{5r^6-2R^6}\right). \label{crucial}
\eeq
 
We see that the spatial components $\overline{\cal H}_{ij}$ have the expected tensor structure. On the other hand, unlike for the T$_4$ modes above, now the $\tau\tau$ component $\overline{\cal H}_{\scalebox{0.9}{$\tau\tau$}}\propto {\cal T}_{\scalebox{0.9}{$\tau\tau$}}$ is nonvanishing. Since  ${\cal T}_{\scalebox{0.9}{$\tau\tau$}}=0$ for the soliton configuration as we  pointed out below (\ref{convert}), naively this implies that the soliton does not excite the S$_4$ mode. However,  the situation is more complex  because of the presence of ${\cal T}^{\rm other}$ in (\ref{other}) and its induced  metric fluctuation $\delta g^{\rm other}$.  The primary role of this term is to account for the ${\cal T}_{\mu \scalebox{0.9}{$r$}}$ and ${\cal T}_{\scalebox{0.9}{$rr$}}$ components of the bulk energy momentum tensor. These  are nonvanishing for the soliton solution (see (\ref{zz})-(\ref{convert})), but they do not directly couple to any of the above TT modes. In the axial gauge $\delta g_{M\scalebox{0.9}{$r$}}=0$, the  components of the Einstein equation having an $r$-index  serve as  first-order  constraints \cite{Friess:2006fk} that can be used to eliminate non-TT modes via the conservation law $\nabla^M{\cal T}_{Mn}=0$ ($n=0,1,2,3,11$), or explicitly,   
\beq
\partial^\mu {\cal T}_{\mu n}&=& -\frac{1}{L^4r^3}\partial_r (r(r^6-R^6){\cal T}_{\scalebox{0.9}{$r$}n}), 
\label{delT}\\
\eta^{mn}{\cal T}_{mn} &=& \frac{1}{L^4r^2}\left(L^4r^3\partial^\mu {\cal T}_{\mu \scalebox{0.9}{$r$}} +r(r^6-R^6)\partial_r {\cal T}_{\scalebox{0.9}{$rr$}} + (8r^6+R^6){\cal T}_{\scalebox{0.9}{$rr$}}\right) 
-\frac{r^6(r^6+2R^6)}{(r^6-R^6)^2}{\cal T}_{\scalebox{0.9}{$\tau\tau$}}.
\label{etaT}
\eeq

Let us 
employ the following ansatz 
\beq
\delta g^{\rm other}_{\mu\nu}(\vec{k},r)&=&  \frac{r^2}{L^2} \frac{k_\mu k_\nu}{k^2} a(r,k),\nn
\delta g^{\rm other}_{11,11}(\vec{k},r)&=&\frac{r^2}{L^2} b(r,k), \label{parameteri}\\
\delta g^{\rm other}_{\scalebox{0.9}{$\tau\tau$}}(\vec{k},r) &=& \frac{r^2f(r)}{L^2} (-b(r,k)). \nonumber
\eeq
This solves the linearized Einstein equation with the bulk energy momentum tensor of the form:
\beq
 2\kappa_7^2  {\cal T}^{\rm other}_{\mu\nu} &\equiv& \frac{1}{L^4r^3}\partial_r \left[\left(\eta_{\mu\nu}-\frac{k_\mu k_\nu}{k^2}\right) r(r^6-R^6)\partial_r a -3R^6\eta_{\mu\nu} b\right], \label{Tother}\\
 2\kappa_7^2  {\cal T}^{\rm other}_{11,11} &\equiv& \frac{1}{L^4r^3}\partial_r \left(r(r^6-R^6) \partial_r(a-b)-3R^6b\right) +k^2 b, \\ 
2\kappa_7^2 {\cal T}^{\rm other}_{\mu \scalebox{0.9}{$r$}}&\equiv& ik_\mu \frac{3R^6}{r(r^6-R^6)}b, \label{mur} \\
2\kappa_7^2 {\cal T}^{\rm other}_{ \scalebox{0.9}{$rr$}} &\equiv& \frac{(5r^6-2R^6)\partial_r a + 3R^6\partial_r b}{r(r^6-R^6)}, \label{trr}\\
2\kappa_7^2 {\cal T}^{\rm other}_{\scalebox{0.9}{$\tau\tau$}}&\equiv& \frac{f(r)}{L^4r^3} \partial_r\left(r(r^6-R^6)\partial_r (a+b) -3R^6(a+b)\right) -f(r) k^2b  . \label{tautau}
\eeq
$a(r,k)$ and $b(r,k)$ are determined by imposing
\beq
{\cal T}^{\rm other}_{ \mu \scalebox{0.9}{$r$}}={\cal T}_{\mu \scalebox{0.9}{$r$}},~~~~
{\cal T}^{\rm other}_{ \scalebox{0.9}{$rr$}}={\cal T}_{ \scalebox{0.9}{$rr$}},
\label{TrrTmur}
\eeq
where the right hand side of these equations is the bulk energy momentum tensor (\ref{convert}). 
Because the gauge configuration under consideration is spherically symmetric, one can show that ${\cal T}_{\mu \scalebox{0.9}{$r$}}$ is proportional to $k_\mu$ and hence the first equation of (\ref{TrrTmur}) can be solved immediately. The second equation of (\ref{TrrTmur})
contains at most first order derivatives in $r$ and it can be uniquely solved by imposing the boundary condition $a\rightarrow 0$ at $r\rightarrow\infty$.
With the conditions (\ref{TrrTmur}), we find that the metric fluctuation (\ref{parameteri}) satisfies the $\mu r$ and $rr$ components of the linearized Einstein equation (\ref{equivalent}).
Note that the above parameterization (\ref{parameteri}) is not unique. Different parameterizations lead to different decompositions of the Einstein equation (\ref{other}) and (\ref{s4t4}) without changing the total induced metric $\delta g_{ab}$ and hence the boundary energy momentum tensor $\langle T_{ab}\rangle$.  We are led to the choice (\ref{parameteri})  because then the reduced 5D energy momentum tensor 
 \beq
 \frac{L}{r}\sqrt{g_{\scalebox{0.9}{$\tau\tau$}}g_{11,11}} {\cal T}^{\rm other}_{\mu\nu} \sim \sqrt{f(r)}\frac{r}{L}   \frac{1}{r^3}\partial_{\scalebox{0.9}{$r$}} [\cdots]= \frac{3}{LR^3} {\rm sgn}(z) \partial_z [\cdots], \label{5d}
 \eeq
is a total derivative in $z$,\footnote{More precisely, because of the sign function ${\rm sgn}(z)=z/|z|$, there arises a delta function contribution $\partial_z {\rm sgn}(z)=2\delta(z)$ upon partial integration. However, this term   vanishes due to the property (\ref{horizon}) of the baryon configuration.} a feature that will turn out to be convenient shortly.   As $r$ goes to infinity, ${\cal T}^{\rm other}_{ \scalebox{0.9}{$rr$}}={\cal T}_{ \scalebox{0.9}{$rr$}}\sim 1/r^{11}$ and ${\cal T}^{\rm other}_{ \mu \scalebox{0.9}{$r$}}={\cal T}_{\mu \scalebox{0.9}{$r$}}\sim 1/r^{10}$ as can be seen by substituting (\ref{sch1}) into  (\ref{zz}) and (\ref{tzz}) and noticing that $G\sim 1/z\sim 1/r^3$ and $H\sim 1/z^2\sim 1/r^6$ in this limit. This implies $a\sim 1/r^9$ and $b\sim 1/r^3$
and hence $\delta g_{\mu\nu}^{\rm other}\sim 1/r^7$ and $\delta g_{11,11}^{\rm other}\sim\delta g_{\scalebox{0.9}{$\tau\tau$}}^{\rm other}\sim 1/r$.
We see that $\delta g_{\mu\nu}^{\rm other}$ decays too fast to contribute to $\langle T_{\mu\nu}\rangle$, but $g_{11,11}^{\rm other}$ and $\delta g_{\scalebox{0.9}{$\tau\tau$}}^{\rm other}$ decay slowly and potentially contribute to $\langle T_{\scalebox{0.9}{$\tau\tau$}}\rangle$ and $\langle T_{11,11}\rangle$\footnote{Note that this part is transverse-traceless by itself $\langle T_{\scalebox{0.9}{$\tau\tau$}}\rangle+\langle T_{11,11}\rangle=0$. It is another TT mode different from the ones described above.} 
(but not to $\langle T_{\mu\nu}\rangle$). 
  On the other hand,   near $z=0$ or $r=R$ where the instanton approximation is good, we can use (\ref{instanton}) to deduce the singular  behavior as $z\to 0$ 
\beq
{\cal T}_{\mu \scalebox{0.9}{$r$}} \sim  \frac{{\cal T}_{\mu z}}{z} \sim  \frac{\partial_z \widehat{A}_0}{z^2}\sim \frac{1}{z}, \qquad {\cal T}_{\scalebox{0.9}{$rr$}} \sim \frac{1}{z^3}.
\eeq
 (The leading singularity comes from the U(1) part.) Comparing with (\ref{mur}) and (\ref{trr}), we find that 
 \beq
 a\sim b\sim \sqrt{r^6-R^6}\sim z. \qquad (z\to 0.)\label{horizon}
 \eeq

Since the $\tau\tau$ component of the total bulk energy momentum tensor ${\cal T}_{ \scalebox{0.9}{$\tau\tau$}}$, as well as
the T$_4$ part ${\cal T}^{{\rm T}(2)}_{ \scalebox{0.9}{$\tau\tau$}}$ and ${\cal T}^{{\rm T}(0)}_{ \scalebox{0.9}{$\tau\tau$}}$, is zero, (\ref{tautau}) means that an effective source term for the S$_4$ mode is induced.
To compensate (\ref{tautau}) by the S$_4$ mode, we impose
\beq
2\kappa_7^2{\cal T}^{\rm S}_{ \scalebox{0.9}{$\tau\tau$}} \equiv\frac{r^2f(r)}{L^2}\frac{5(r^6-R^6)}{5r^6-2R^6}{\cal H}^{\rm S}= -2\kappa_7^2{\cal T}^{\rm other}_{  \scalebox{0.9}{$\tau\tau$}},  \label{tautauS}
\eeq
(see (\ref{s4metH})), which implies
\beq
 2\kappa_7^2 {\cal T}^{\rm S}_{\mu\nu}
 &\equiv&
 -\frac{r^2}{L^2} \frac{5(r^6+2R^6)}{4(5r^6-2R^6)}\left(\eta_{\mu\nu}-\frac{k_\mu k_\nu}{k^2}\right) {\cal H}^{\rm S}\nn 
 &=& \frac{r^6+2R^6}{4f(r)(r^6-R^6)}\left(\eta_{\mu\nu}-\frac{k_\mu k_\nu}{k^2}\right) 2\kappa_7^2 {\cal T}_{\scalebox{0.9}{$\tau\tau$}}^{\rm other} \label{tmunus} \\ 
 &=& \frac{r^6+2R^6}{4(r^6-R^6)}\left(\eta_{\mu\nu}-\frac{k_\mu k_\nu}{k^2}\right) \left[ \frac{1}{L^4r^3} \partial_r\left(r(r^6-R^6)\partial_r (a+b) -3R^6(a+b)\right) - k^2b \right], \nn
\delta g_{\mu\nu}^{\rm S} &\approx&  \frac{r^2}{4L^2}\left(\eta_{\mu\nu}-\frac{k_\mu k_\nu}{k^2}\right)h^{\rm S} + {\cal O}(1/r^7),  \label{gS}
\eeq
with ${\cal H}^{\rm S}\sim 1/r^3$ and $h^{\rm S}\sim 1/r^3$ as $r\to \infty$.  Comparing the large-$r$ behavior of  (\ref{tautauS})
\beq
\frac{r^2}{L^2}{\cal H}^{\rm S} \sim -\frac{1}{L^4r^3} \partial_r (r^7\partial_r b) + k^2 b,
\eeq
with that of (\ref{extra}), we can identify 
\beq
 h^{\rm S}=-b + {\cal O}(1/r^6).  \label{hlarge}
\eeq
(Note that $\partial_r (r^7 \partial_r r^{-6})=0$.)

 We thus conclude that, contrary to the naive expectation, the S$_4$ glueballs in general contribute to the boundary energy momentum tensor, albeit in a somewhat indirect way. 
Consequently, the  source term for the T$_4$ glueballs is not the total ${\cal T}_{\mu\nu}$ (\ref{use}), but instead\footnote{Since the $rr$, $r\mu$ and $\tau\tau$ components of 
${\cal T}_{AB}^{\rm T}\equiv {\cal T}_{AB} - {\cal T}^{\rm S}_{AB}-{\cal T}^{\rm other}_{AB}$ are all zero,
the conservation law (\ref{delT}) and (\ref{etaT}) imply that
${\cal T}_{AB}^{\rm T}$ satisfy the transverse traceless conditions $\partial^\mu {\cal T}^{\rm T}_{\mu n}=0=\eta^{mn}{\cal T}^{\rm T}_{mn}$ $(n,m=0,1,2,3,11)$ in 5 dimensions.
}
\beq
{\cal T}_{\mu\nu}^{\rm T}\equiv{\cal T}^{\rm T(0)}_{\mu\nu}+{\cal T}^{\rm T(2)}_{\mu\nu}= {\cal T}_{\mu\nu} - {\cal T}^{\rm S}_{\mu\nu}-{\cal T}^{\rm other}_{\mu\nu}. \label{effT}
\eeq
Note that the leading term $\delta g_{\mu\nu}^{\rm S}\sim r^2 h^{\rm S}\sim 1/r$ as $r\to \infty$ 
is canceled against a like term in $\delta g_{\mu\nu}^{\rm T}\propto {\cal T}^{\rm T}_{\mu\nu}\sim -{\cal T}_{\mu\nu}^{\rm S}$. Indeed, from 
 (\ref{s4metH}) we find 
\beq
&& 2\kappa_7^2 {\cal T}_{\mu\nu}^{\rm S} 
\approx  \left(-\frac{1}{L^4r^3} \partial_r (r^7-rR^6)\partial_r +k^2\right) \frac{L^2}{r^2}\delta g_{\mu\nu}^{{\rm S}}. 
\eeq
Similarly, from (\ref{00s}) and (\ref{0+g}) 
\beq
2\kappa_7^2{\cal T}_{\mu\nu}^{{\rm T}} 
= \left(-\frac{1}{L^4r^3} \partial_r (r^7-rR^6)\partial_r +k^2\right) \frac{L^2}{r^2}\delta g_{\mu\nu}^{{\rm T}}.
\eeq
Therefore, near the boundary  $r\to \infty$,
\beq
2\kappa_7^2 \left( {\cal T}_{\mu\nu}^{{\rm T}}+{\cal T}_{\mu\nu}^{\rm S}\right) 
\approx \left(-\frac{1}{L^4r^3} \partial_r (r^7-rR^6)\partial_r +k^2\right) \frac{L^2}{r^2}\left( \delta g_{\mu\nu}^{{\rm T}}+\delta g_{\mu\nu}^{{\rm S}} \right) . \label{diff}
 \eeq
 Since ${\cal T}_{\mu\nu}^{{\rm T}}+{\cal T}_{\mu\nu}^{\rm S} = {\cal T}_{\mu\nu}-{\cal T}_{\mu\nu}^{\rm other}\sim   1/r^7$, most of the terms in $\delta g_{\mu\nu}^{\rm T/S}$ which decay slower than $1/r^7$  cancel in the sum $\delta g_{\mu\nu}^{{\rm T}}+\delta g_{\mu\nu}^{{\rm S}}$. The leading surviving contribution is the  $1/r^4$ term 
 \beq
 \delta g_{\mu\nu}^{{\rm T}}+\delta g_{\mu\nu}^{{\rm S}} \sim \frac{C_{\mu\nu}}{r^4}. \qquad (r\to \infty)
 \label{ggC}
 \eeq
 This is precisely what contributes to the boundary energy momentum tensor via holographic renormalization (\ref{holoreno}). Remarkably,  when $k=0$, $C_{\mu\nu}$ is directly proportional to the classical energy momentum tensor (\ref{return}) as can be seen by writing 
\beq
2\kappa_7^2 \int_R^\infty dr r^3 ({\cal T}_{\mu\nu}-{\cal T}_{\mu\nu}^{\rm other})  = 2\kappa_7^2\int_R^\infty dr r^3 ({\cal T}^{\rm T}_{\mu\nu}+{\cal T}_{\mu\nu}^{\rm S}) = \frac{6C_{\mu\nu}}{L^2},
\eeq
at $k=0$. In the last equality, we have used the identity (\ref{crucial}) in order to recast the integrand (cf. (\ref{tmunus})) as a total derivative. 
The first term on the left hand side is proportional to  (\ref{return}) and the second term vanishes (cf. (\ref{horizon}))
\beq
2\kappa_7^2\int_R^\infty dr r^3 {\cal T}_{\mu\nu}^{\rm other} =  \frac{3R^6}{L^4}\eta_{\mu\nu}b|_{r=R} =0. \label{r=R}
\eeq
 This means that the naive expression (\ref{return}) can be used to calculate gravitational form factors  at $k=0$. 
 Below we shall reconfirm this important observation in a different way.

 \subsection{Glueball dominance of the gravitational form factors}
 
Armed with the general discussion in the previous subsection, we are now ready to compute the form factor $D(\vec{k})$ for generic values of $\vec{k}$. From (\ref{cancel}), we can write 
\beq
D= D^{{\rm T}(2)}+D^{{\rm T}(0)} + D^{{\rm S}(0)} =4M( t_2+t_0+s_0), 
\eeq
where each component can be determined by solving the corresponding wave equations. 
It should be clear by now that the split $D^{\rm T}\sim t_2+t_0$ is actually not necessary.  As we have seen above, the 
propagators of the T$_4(2^{++})$ and T$_4(0^{++})$ modes are identical and diagonal in Lorentz indices 
\beq
G^{{\rm T}(2)}_{\mu\nu AB} =G^{{\rm T}(0)}_{\mu\nu AB} \equiv  \frac{r^2r'^2}{L^4}\tilde{G}^{\rm T}(\eta_{\mu \rho}\eta_{\nu\lambda}+\eta_{\mu\lambda}\eta_{\nu\rho}) \delta^\rho_A \delta^\lambda_B, \label{t4eq}
\eeq
where $\tilde{G}^{\rm T}$ is the Green function for the massless Klein-Gordon equation 
\beq
-\nabla^2 \tilde{G}^{\rm T}(r,r',\vec{k}) = -\left(\frac{1}{L^2r^5}\partial_r \Bigl((r^7-rR^6)\partial_r\Bigr)- \frac{L^2}{r^2}\vec{k}^2 \right)\tilde{G}^{\rm T}(r,r',\vec{k}) = \frac{L^5}{r^5}\delta(r-r').
\label{sol}
\eeq
We can thus write, in (\ref{space}), 
\beq
G^{{\rm T}(2)}_{\mu\nu AB}{\cal T}^{AB}_{{\rm T}(2)} +G^{{\rm T}(0)}_{\mu\nu AB}{\cal T}^{AB}_{{\rm T}(0)}= \frac{r^2r'^2}{L^4}\tilde{G}^{\rm T}(\eta_{\mu\rho}\eta_{\nu\lambda}+\eta_{\mu\lambda}\eta_{\nu\rho}){\cal T}^{\rho\lambda}_{\rm T}  = \frac{2r^2}{r'^2}\tilde{G}^{\rm T}{\cal T}^{\rm T}_{\mu\nu}, \label{spatial}
\eeq 
where ${\cal T}^{\rm T}_{\mu\nu}$ is given by (\ref{effT}). This leads to the formula 
\beq
 \delta g^{\rm T}_{\mu\nu}(r,\vec{k})&=&\! \! 
 2\kappa_7^2 \frac{r^2}{L^2} \int^\infty_R dr' \sqrt{-G'^{(7)}}\tilde{G}^{\rm T}(r,r',\vec{k})\frac{L^2}{r'^2}{\cal T}^{\rm T}_{\mu\nu}(r',\vec{k}) \nn 
&=&\! \!2\kappa_7^2 
\frac{\pi^2Cr^2L^2}{3g_s}\int_R^\infty \frac{dr'}{\sqrt{f(r')}}\tilde{G}^{\rm T}(r,r',\vec{k})  {\rm tr}[F_{\mu\rho}F_{\nu}^{\ \rho}(r',\vec{k})] + \cdots \label{final} \\
&=& \!\!\kappa_7^2\frac{\pi^2Cr^2L^2R}{9g_s}\int_{-\infty}^\infty dz \tilde{G}^{\rm T}(r,r'(z),\vec{k})h(z) {\rm tr}[ F_{\mu\rho}F_{\nu}^{\ \rho}(r'(z),\vec{k})] + \cdots, \nonumber
\eeq
where we used (\ref{use}) and  in the second line we switched to the variable $z$ using (\ref{z}). The omitted terms include the subtraction $-{\cal T}_{ij}^{\rm S}-{\cal T}_{ij}^{\rm other}$ in (\ref{effT}).
Similarly, with the help of the identity (\ref{crucial}), we introduce the Green function for the S$_4$ mode
\beq
-\left[\frac{1}{L^2r^5}\partial_r \left((r^7-rR^6)\left(\partial_r + \frac{144r^5R^{6}}{(5r^6-2R^6)(r^6+2R^6)}\right)\right) -\frac{L^2}{r^2}\vec{k}^2\right]\tilde{G}^{\rm S}(r,r',\vec{k}) = \frac{L^5}{r^5}\delta(r-r'). \label{greS}
\eeq
Using (\ref{s4met}) and (\ref{s4metH}), we find 
\beq
\delta g_{\mu\nu}^{\rm S}(r,\vec{k})\approx  2\kappa_7^2 \frac{r^2}{L^2}\int_R^\infty dr'\sqrt{-G'_{(7)}}\tilde{G}^{\rm S}(r,r',\vec{k}) \frac{L^2}{r'^2}{\cal T}_{\mu\nu}^{\rm S}(r',\vec{k}) , \label{write}
\eeq
up to terms  irrelevant (suppressed by powers of $1/r$) to holographic renormalization.  

There is, however, a subtlety in the above discussion. As we have pointed out in the previous subsection, both $\delta g^{\rm T}_{\mu\nu}$ and $\delta g^{\rm S}_{\mu\nu}$ have a component which decays slowly $\sim 1/r$. While they cancel eventually in the sum $\delta g^{\rm T}_{\mu\nu}+\delta g^{\rm S}_{\mu\nu}$, it is convenient to extract the ${\cal O}(1/r^4)$ components in $\delta g^{\rm T/S}_{\mu\nu}$ by subtracting the slowly decaying terms, because each term in our formal expression of the Green functions $\tilde{G}^{\rm T/S}(r,r')$ in (\ref{sum}) behaves as $1/r^6$ as $r\rightarrow\infty$ and the $1/r$ terms in $\delta g_{\mu\nu}^{\rm T/S}$ cannot be captured without taking the infinite sum.
For this purpose, we define
\beq
\overline{\delta g}_{\mu\nu}^{\rm T} \equiv \delta g_{\mu\nu}^{\rm T} -\frac{r^2}{4L^2} \left(\eta_{\mu\nu}-\frac{k_\mu k_\nu}{k^2}\right)b, \nn
\overline{\delta g}_{\mu\nu}^{\rm S}\equiv  \delta g_{\mu\nu}^{\rm S} + \frac{r^2}{4L^2} \left(\eta_{\mu\nu}-\frac{k_\mu k_\nu}{k^2}\right)b.
\eeq
and  the corresponding subtracted bulk energy momentum tensor 
\beq
2\kappa_7^2\overline{{\cal T}}^{\rm T}_{\mu\nu}&\equiv&  2\kappa_7^2{\cal T}^{\rm T}_{\mu\nu} + \frac{1}{4}\left(\eta_{\mu\nu}-\frac{k_\mu k_\nu}{k^2}\right)\left(-k^2+\frac{1}{L^4r^3}\partial_r \left((r^7-rR^6)\partial_r\right) \right)b, \nn
 2\kappa_7^2\overline{{\cal T}}_{\mu\nu}^{\rm S}&\equiv& 2\kappa_7^2{\cal T}^{\rm S}_{\mu\nu} - \frac{1}{4}\left(\eta_{\mu\nu}-\frac{k_\mu k_\nu}{k^2}\right) \nn 
&& \times \left[-k^2+\frac{1}{L^4r^3}\partial_r\left( (r^7-rR^6)\left(\partial_r +\frac{144r^5R^6}{(5r^6-2R^6)(r^6+2R^6)}\right)\right)\right]b.
\eeq
(\ref{gS}), (\ref{hlarge}) and (\ref{ggC}) imply
$\overline{\delta g}_{\mu\nu}^{\rm S/T}\sim 1/r^4$ as desired.
 
With this in mind, we now solve (\ref{sol}) and (\ref{greS}) by introducing the complete set of eigenfunctions 
\beq
&&-\frac{1}{r^3L^4}\partial_r \Bigl((r^7-rR^6)\partial_r \Psi^{\rm T}_n(r)\Bigr)=(m^{\rm T}_n)^2 \Psi^{\rm T}_n(r), \qquad (n=1,2,3,\cdots) \label{complete} \\
&& -\frac{1}{r^3L^4} 
\partial_r\left((r^7-rR^6)\left(\partial_r + \frac{144r^5R^6}{(5r^6-2R^6)(r^6+2R^6)}\right)\Psi^{\rm S}_n(r)\right)=(m^{\rm S}_n)^2 \Psi^{\rm S}_n(r), \label{msmass}
\eeq
which are regular at $r=R$ and vanish in the limit $r\to \infty$. 
They satisfy the completeness relations 
\beq
\frac{r^3}{L^3} \sum_{n=1}^\infty \Psi^{\rm T/S}_n(r)\Psi^{\rm T/S}_n(r')=\delta(r-r'),
\label{completeness}\\
\int_R^\infty dr r^3 
\Psi^{\rm T/S}_n(r)\Psi^{\rm T/S}_m(r)=L^3 \delta_{mn}.
\eeq 
The solution of (\ref{sol}) and (\ref{greS}) can then be formally written as 
\beq
\tilde{G}^{\rm T/S}(r,r',\vec{k}) =\sum_{n=1}^\infty \frac{\Psi^{\rm T/S}_n(r)\Psi^{\rm T/S}_n(r')}{(m_n^{\rm T/S})^2+\vec{k}^2}. \label{sum}
\eeq
$m^{\rm T/S}_n$ are nothing but the masses of the T$_4$ and S$_4$ 0$^{++}$ glueballs  \cite{Brower:2000rp}. Numerically 
(see Table 1 of \cite{Brower:2000rp}), 
\beq
\frac{L^4}{R^2}(m^{\rm T}_{n})^2= 22.097, \quad 55.584,\quad 102.456, \quad 162.722, \quad 236.400, \quad 323.541, \quad 424.195,\cdots
\eeq
or equivalently, using the relation $L^2/R=3/M_{KK}$, 
\beq
\frac{m^{\rm T}_{n}}{M_{KK}}=1.567,\quad  2.485,\quad 3.374,\quad 4.242,\quad 5.125,\quad 5.996, \quad 6.865,\cdots.
\label{2mass}
\eeq 
Surprisingly, 
to a very good approximation,
\beq
m^{\rm T}_{n}\approx 2m_n,
\eeq
where $m_{n}$ are the  vector meson masses (\ref{masses}). This approximate relation becomes more and more accurate as $n$ increases. Already when  $n=2$, the discrepancy is at the sub-percent level. This  hints at some kind of degeneracy between a glueball and a pair of  vector  mesons.   We shall come back to this point in the concluding section. Similarly, for the S$_4$ mode, 
\beq
\frac{m^{\rm S}_{n}}{M_{KK}} = 0.901, \quad 2.285, \quad 3.240, \quad 4.150, \quad 5.042, \quad 5.925, \quad 6.804,\cdots. \label{s4masses}
\eeq  

 It is easy to see that the normalizable modes $\Psi^{\rm T/S}_n(r)$ decay near the boundary $r\to \infty$ as 
\beq
\Psi^{\rm T/S}_n(r)\approx \frac{\alpha^{\rm T/S}_n}{r^6}.
\eeq
The coefficients  
$\alpha^{\rm T/S}_n$ can be obtained by integrating (\ref{complete}) and (\ref{msmass}) over $r$ 
\beq
\alpha^{\rm T/S}_n=\frac{(m^{\rm T/S}_{n})^2L^4}{6}\int_R^\infty dr r^3 \Psi^{\rm T/S}_n(r). \label{alphan}
\eeq
We can thus write the leading term as
\beq
\overline{\delta g}_{\mu\nu}^{\rm T/S}(\vec{k},r)\approx  \frac{C^{\rm T/S}_{\mu\nu}(\vec{k})}{r^4}+\cdots
\eeq
with $\vec k$-dependent coefficients $C_{\mu\nu}^{T/S}(\vec k)$.
  The relation between $\langle T_{\mu\nu}\rangle$ and $C^{\rm T/S}_{\mu\nu}$ is fixed by  holographic renormalization \cite{deHaro:2000vlm,Kanitscheider:2008kd}. This is conveniently done in a different coordinate system
  \beq
  \varrho=\frac{L^4}{r^2\left(\frac{1}{2}(1+\sqrt{f(r)})\right)^{2/3}},
  \eeq
  in which the metric (\ref{fe}) takes the Fefferman-Graham form 
\beq
ds^2 &=& L^2\left[\frac{1}{\varrho}\left(1-\frac{5\varrho^3}{6\varrho_{KK}^3}\right) d\tau^2 +\frac{1}{\varrho}\left(\left(1+\frac{\varrho^3}{6\varrho_{KK}^3}\right) dx^\mu dx_\mu + \frac{\varrho^3}{L^{10}}C_{ij}dx^i dx^j\right) + \frac{d\varrho^2}{4\varrho^2} + \frac{1}{4}d\Omega_4^2 \right] \nn
&& + {\cal O}(\varrho^3), 
\eeq
where $\varrho_{KK}=L^4/R^2$. 
From this expression we can read off the boundary energy momentum tensor 
\beq
&&\langle T_{\mu\nu}(\vec{k})\rangle =
\langle T^{\rm T}_{\mu\nu}(\vec{k})\rangle +
\langle T^{\rm S}_{\mu\nu}(\vec{k})\rangle,
\label{tmunuk}
\\
&&\langle T^{\rm T/S}_{\mu\nu}(\vec{k})\rangle = \frac{6}{2\kappa^2_{7}L^{5}}C^{\rm T/S}_{\mu\nu}(\vec{k})= \frac{6}{L^{10}} \sum_{n=1}^\infty \alpha_n^{\rm T/S} \int_R^\infty dr' \frac{r'^3 \Psi_n^{\rm T/S}(r')}{\vec{k}^2+(m_n^{\rm T/S})^2} \overline{\cal T}^{\rm T/S}_{\mu\nu}(r',\vec{k}),
\label{tmunuTS}
\eeq
where $\alpha^{\rm T/S}_n$ are given by (\ref{alphan}). In particular, using (\ref{kappa1}) and (\ref{useful}), we find 
\beq
\langle T^{\rm T}_{ij}(\vec{k})\rangle = \frac{12\kappa}{L^7} \sum_n \frac{\alpha^{\rm T}_n }{\vec{k}^2+(m^{\rm T}_{n})^2} \int_{-\infty}^\infty dz \Psi^{\rm T}_n(r'(z))  h(z){\rm tr}\left[F_{ik}F_{jk}(z,\vec{k})\right]+\cdots.
\label{holo}
\eeq
To get the four-dimensional energy momentum tensor, we have to integrate over the extra dimensions $x_{11}$ and $\tau$, but this has been already done in (\ref{space}).

We have thus seen that the D-term gravitational form factor is saturated by the exchange of an infinite tower of $2^{++}$ and $0^{++}$  glueballs. Schematically, 
\beq
D(|\vec{k}|) \sim \sum_{n=1}^\infty \frac{c^{\rm T}_n(|\vec{k}|)}{\vec{k}^2+(m^{\rm T}_{n})^2} + \sum_{n=1}^\infty \frac{c^{\rm S}_n(|\vec{k}|)}{\vec{k}^2+(m^{\rm S}_{n})^2} . \label{single}
\eeq
This should be contrasted with the tripole form 
\beq
D(|\vec{k}|)\sim \frac{1}{(\vec{k}^2+\Lambda^2)^3}, \label{tri}
\eeq
 suggested by perturbative QCD analyses at large-$k$ \cite{Tanaka:2018wea,Tong:2021ctu,Tong:2022zax}.
 While (\ref{single}) and (\ref{tri}) are seemingly very different, an infinite sum can change the analytic properties in $|\vec{k}|$. Indeed, it has been observed in \cite{Hashimoto:2008zw} that the electromagnetic form factors calculated in the present model can be well fitted by the dipole form $1/(\vec{k}^2+\Lambda'^2)^2$ despite being formally expressed by an infinite sum of single pole (vector meson) propagators like (\ref{single}). It is tempting to refer to  (\ref{single}) as the  `glueball dominance' of the D-term, in perfect  analogy to the vector meson dominance of the electromagnetic form factors. In both cases, in general one has to sum over infinitely many resonances.

The summation over $n$ cannot be performed in a closed analytic form. However,   the value at the special point $k=0$ can be exactly determined with the help of  (\ref{alphan})
\beq
\lim_{\scalebox{0.9}{$r$}\to \infty} \tilde{G}^{\rm T/S}(r,r',\vec{k}=0)= \frac{1}{r^6}\sum_{n} \frac{\alpha^{\rm T/S}_n\Psi^{\rm T/S}_n(r') }{(m^{\rm T/S}_{n})^2} = \frac{L^4}{6r^6}\sum_{n}\int_R^\infty dr''r''^3 \Psi^{\rm T/S}_n(r'')\Psi^{\rm T/S}_n(r') =\frac{L^7}{6r^6} .\label{k0}
\eeq
Remarkably, this is independent of $r'$, so the convolution  with the graviton propagator  disappears.   
Thus,  the distinction between the S$_4$ and T$_4$ propagators become irrelevant at this point.   
It then follows  that 
\beq
\langle T_{ij}(\vec{k}=0) \rangle \approx 2\kappa \int dz h(z){\rm tr}[F_{il}F_{jl}](\vec{k}=0,z)+\cdots. \label{naive2}
\eeq
This is nothing but the  naive classical formula (\ref{zint}) evaluated in the previous section except for the subtraction term $-{\cal T}^{\rm other}$ in (\ref{effT}) which, however, can be ignored because it is a total derivative (\ref{r=R}).   We have thus arrived at the same conclusion as in the previous subsection that  the value $D(0)$ can be calculated by simply Fourier transforming the classical energy momentum tensor (\ref{zint}) without any reference to glueballs.   
 At $\vec{k}=0$, exchanging infinitely many glueballs is tantamount to exchanging no glueball at all.

Another regime where an analytical expression of the propagator is available is the large momentum region $|\vec{k}|\gg M_{KK}$.\footnote{The following discussion should be taken with caution because the present model starts to deviate from QCD as the momentum transfer gets much larger than $M_{KK}$.} In this regime, we expect that  $\tilde{G}^{\rm T/S}(r\to \infty ,r',\vec{k})$  is exponentially small in $|\vec{k}|$ for $|\vec{k}|L^2 > r'$.\footnote{When $|\vec{k}|L^2\gg  r'\gg R$, one can see from (\ref{largek}) that the propagator is exponentially suppressed, and the suppression gets  stronger as $r'$ approaches $R$ from above. When $r'\sim R$, the AdS approximation breaks down. Nevertheless, by formally Taylor expanding (\ref{sum}) 
\beq
\tilde{G}^{\rm T/S}(r,r',\vec{k})=\sum_{n=1}^\infty \sum_{i=1}^\infty \frac{(-(m^{\rm T/S}_{n})^2)^{i-1}}{(\vec{k}^2)^{i}}\Psi^{\rm T/S}_n(r)\Psi^{\rm T/S}_n(r'),
\eeq
and repeatedly using (\ref{complete}), (\ref{msmass}) and (\ref{completeness}), one can show that all the coefficients of $1/(\vec{k}^2)^i$ vanish. This suggests that the propagator decays faster than any inverse power of $1/\vec{k}^2$. 
}
In the opposite region 
\beq
r'> |\vec{k}|L^2=3R\frac{|\vec{k}|}{M_{KK}} \gg R,
\eeq
 the geometry is approximately AdS$_7$ (instead of AdS$_7$ black hole), and the propagator is explicitly known in terms of the modified Bessel functions \cite{Mueck:1998wkz}
\beq
\tilde{G}^{\rm T/S}(r,r',\vec{k}) \approx \frac{L^{7}}{r^3r'^3}K_3(|\vec{k}|L^2/r')I_3(|\vec{k}|L^2/r) \approx \frac{|\vec{k}|^3L^{13}}{48r^6r'^3}K_3(|\vec{k}|L^2/r'), \label{largek}
\eeq
for $r\to \infty >r'$. (\ref{largek}) reduces to (\ref{k0}) by formally setting $\vec{k}=0$. Noting that $r'\approx |z|^{1/3}R$ in this regime, 
we see that the D-term can be calculated from the convolution 
\beq
\langle T_{ij}(\vec{k}) \rangle &\approx&  2\kappa \int dz \left\{\frac{27|\vec{k}|^3}{8|z|M_{KK}^3}K_3\left(\frac{3|\vec{k}|}{|z|^{1/3}M_{KK}}\right)\right\} h(z){\rm tr}[F_{il}F_{jl}](\vec{k},z)+\cdots.
\label{lak}
\eeq
Since the $z$-integral is effectively limited to $|z|\gtrsim 27|\vec{k}|^3/M_{KK}^3$, we see that the gravity effect strongly suppresses the D-term at large-$|\vec{k}|$.

To summarize, we arrive at the following strategy to calculate the energy momentum tensor and the gravitational form factors. The value at the special point $\vec{k}=0$ can be obtained  `classically'  as demonstrated in the previous section. When $0<|\vec{k}|\lesssim M_{KK}\sim 1$ GeV, they can be calculated, in principle, by (\ref{tmunuk}) and (\ref{tmunuTS}) where one has to explicitly evaluate the Fourier modes of the bulk energy momentum tensor ${\cal T}_{\mu\nu}(r,\vec k)$ and perform the infinite sum. Eq.~(\ref{tmunuTS}) exhibits   the phenomenon of `glueball dominance', namely the form factor is saturated by the exchange of the T$_4$ and S$_4$ glueballs. When $|\vec{k}|\gg M_{KK}$, we may continue to use holographic renormalization. 
The bulk spacetime effectively reduces to AdS$_7$ (as opposed to the AdS$_7$ black hole) and the connection to glueballs becomes less obvious. 
Note that the first region $|\vec{k}|\ll 1$ GeV practically covers the phenomenologically important region \cite{Hatta:2018ina,Hatta:2019lxo,Guo:2021ibg}. The region $|\vec{k}|\gg 1$ GeV is also interesting for certain applications  \cite{Boussarie:2020vmu,Hatta:2021can,Sun:2021gmi}, but in this regime one should use perturbative QCD techniques    \cite{Tong:2021ctu,Tong:2022zax} rather than holography. 


Finally, let us  comment on the relation to previous approaches in the literature \cite{Hatta:2018ina,Mamo:2019mka,Mamo:2021krl}. In bottom-up holographic models, one works in asymptotically AdS$_5$ spaces. Needless to say, the S$_4$ mode is nonexistent in these spaces.  There are five transverse-traceless graviton polarizations in AdS$_5$, in one-to-one correspondence with the T$_4(2^{++})$  modes in AdS$_7$. They are interpreted as the 2$^{++}$ glueballs in the boundary field theory. Via holographic renormalization, the AdS$_5$ counterpart of  (\ref{t4})  induces the transverse-traceless part of the QCD energy momentum tensor, namely, the first line of   (\ref{can}) \cite{Hatta:2018ina}. Separately to this, one has to exchange the dilaton dual to the $0^{++}$ glueballs which leads to the trace part of the energy momentum tensor,  the second line of (\ref{can}). The $A$-form factor in the coefficient of $k^\mu k^\nu-\eta^{\mu\nu}k^2$ cancels between the two, and only the $D$-form factor  remains. However, the split (\ref{can}) is somewhat artificial, and the cancellation may become a delicate issue \cite{Mamo:2021krl}.  In our  AdS$_7$/CFT$_6$/QCD$_4$ setup, the $2^{++}$ and $0^{++}$  glueballs (the latter would correspond to the dilaton in a five-dimensional setup) are treated on an equal footing.  They are both transverse-traceless  from  seven-dimensional point of view, and their distinction is immaterial for the purpose of calculating the gravitational form factors.

\section{Discussions and conclusions}

\begin{figure}
  \includegraphics[width=1\linewidth]{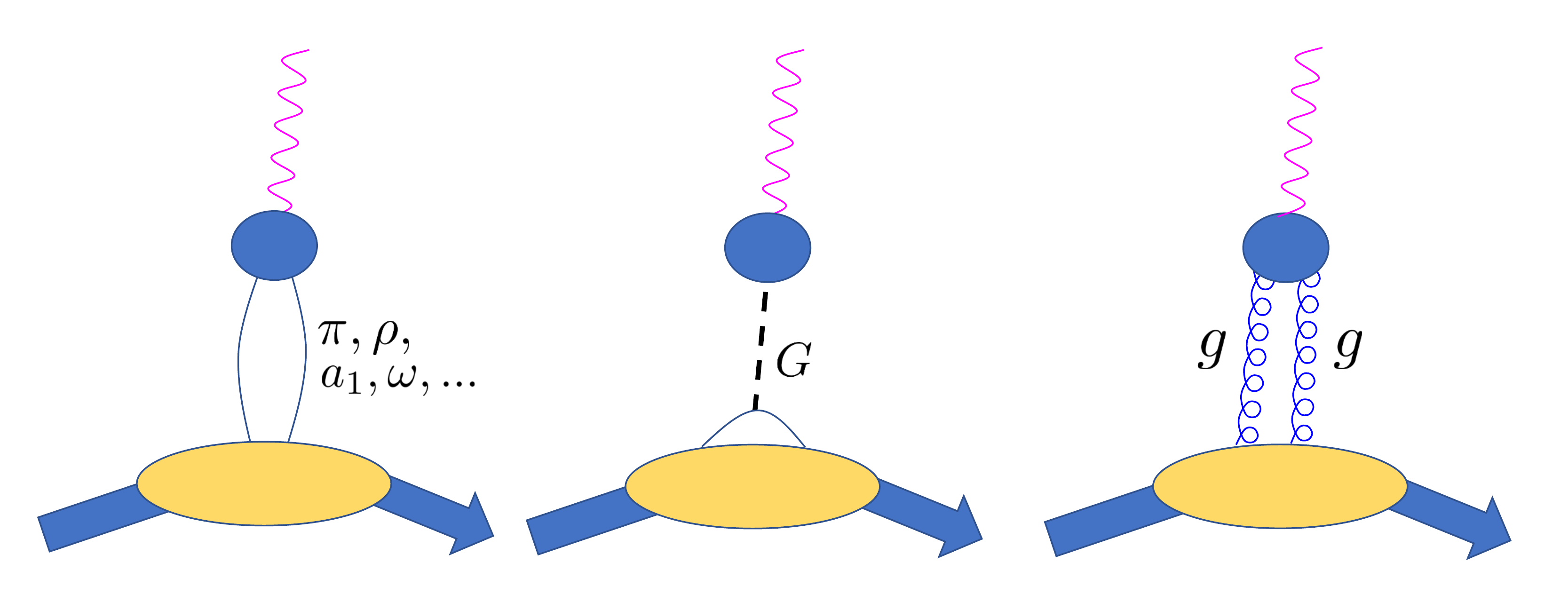}
\caption{Diagrammatic interpretation of the nucleon's  gravitational form factor $D(|\vec{k}|)$. The wavy line denotes a graviton.  Left: two vector meson (pion) exchanges at $\vec{k}=0$; Middle: $2^{++}$ and $0^{++}$  glueball exchanges when $|\vec{k}|\lesssim 1$ GeV; Right: two gluon exchanges in the perturbative high-$|\vec{k}|$ ($\gg 1$ GeV) region. 
}
\label{1}
\end{figure}

Let us try to interpret our results diagrammatically. In our model, the form factor  $D(t=-\vec{k}^2)$ is given by the convolution of the soliton energy momentum tensor $T^{AB}$  and the graviton propagator in the AdS$_7$ black hole geometry which contains information about the $2^{++}$ and $0^{++}$ glueballs. When $t=0$, the graviton propagator trivializes. The D-term $D(0)$ is then related to the  Fourier transform of the `classical' energy momentum tensor $T^{cl}$  (\ref{return}) evaluated at on-shell. Since $T^{cl}\sim G^2,H^2$ (see (\ref{tedious})) can be expressed by  the square of vector meson propagators near the boundary, one may say that a graviton couples to the nucleon by exchanging pairs of pions and infinitely many  vector and axial-vector mesons.\footnote{The energy momentum tensor also contains cubic and quartic terms in $G$ and $H$ (c.f. (\ref{tedious})) which can be interpreted as three-meson and four-meson exchanges. However, such contributions are suppressed near the boundary. }   This is depicted in the left diagram of Fig.~\ref{1}. Roughly the D-term may be understood as the coupling between a graviton and a `meson cloud', similarly to the pion cloud in the chiral quark soliton model (see, e.g., \cite{Goeke:2007fp}). We however emphasize that in our calculation the pion contribution  is just one term $n=0$ in the infinite sum (\ref{mesons}).   When $0<|\vec{k}|\lesssim M_{KK}\sim 1$ GeV, one sees the individual contributions of glueballs (\ref{sum}). The emerging picture in this regime is that the  meson pairs from ${\cal T}^{AB}$  in the bulk couple with  glueball intermediate states which eventually interact with  an external graviton (Fig.~\ref{1}, middle). One may interpret this situation as the `glueball dominance' of the gravitational form factors, in perfect analogy to the vector meson dominance  of the electromagnetic form factors previously observed in our model \cite{Hashimoto:2008zw}. In both cases, one has to exchange infinitely many resonances.  Curiously, as we pointed out below (\ref{masses}), the masses of the T$_4$ glueballs are strikingly close to those of two vector mesons. To our knowledge, this coincidence does not seem to have been pointed out in the literature.  It would be very interesting to explore the reason and implications of this observation. Finally, in the large-$k$ region, the geometry becomes effectively AdS$_7$ (as opposed to the AdS$_7$ black hole) and the information about the glueball spectrum is lost. While the meson exchange picture is still possible, in this region   one should rather use perturbative QCD techniques to calculate  form factors. In particular,   the D-term is given by the two gluon exchange to leading order (Fig.~\ref{1}, right) \cite{Tong:2021ctu}.

The connection between the D-term and glueballs has been recently emphasized by Mamo and Zahed using  a bottom-up AdS$_5$/QCD$_4$ model  \cite{Mamo:2021krl,Mamo:2022eui}. As we commented at the end of Section 5, in their calculation the D-term results from a cancellation between the transverse-traceless (TT) graviton and dilaton exchanges in AdS$_5$ (in other words,  between the first and second lines in (\ref{can})). They argue that, due to the degeneracy of the $2^{++}$ and $0^{++}$ glueball masses, the cancellation is complete (meaning $D=0$) in the large-$N_c$ limit. They further argue that the D-term becomes nonzero only after including $1/N_c$ corrections, suggesting that  the large-$N_c$ counting argument \cite{Polyakov:2018zvc}
\beq
D \sim O(N_c^2), \label{largenc}
\eeq
does not hold. 
In contrast, in our  AdS$_7$/CFT$_6$/QCD$_4$ calculation, the contributions from the  $2^{++}$ and $0^{++}$ T$_4$ glueballs add up (see (\ref{t4eq}), (\ref{spatial})),  
after which their distinction becomes immaterial.  Moreover,  the value $D(k=0)$ can be calculated `classically' as in Section 4 without any reference to glueballs. Our result is consistent with the large-$N_c$ counting since 
\beq
D\sim \kappa M  \sim O(N_c^2).
\eeq
 ($M_{KK}\sim O(N_c^0)$ since  it is related to the $\rho$-meson mass.)  Interestingly, however, the actual  numerical value calculated in Section 4 is small due to a cancellation between the iso-singlet and iso-vector contributions.

In conclusion, we have demonstrated that our  model of QCD offers a novel holographic framework to calculate the nucleon D-term and possibly other gravitational form factors, with a vivid physical interpretation in terms of meson and glueball exchanges. There are a number of avenues for improvement and generalization.  First, the quantization of the collective coordinates should be implemented. It is known that the  quantization  brings in uncertainties in baryon masses associated with the zero-point (vacuum) energy of harmonic oscillators \cite{Hata:2007mb}. We expect that the D-term, being a nonforward matrix element, is less affected by this problem. 
However, without quantization, the spin-1/2 nature of the nucleon is lost, and one cannot talk about the $B$-form factor (\ref{gff}). Second, in order to  calculate $D(k)$ for nonvanishing values of $k$, it is desired to convolute a more accurate soliton configuration in the entire space $0<|z|<\infty$ obtained numerically (see e.g.,  \cite{Bolognesi:2013nja,Rozali:2013fna,Suganuma:2020jng})  with the graviton propagator to incorporate the contributions from the glueballs. This is  crucial, in particular, for the calculation  of the `mechanical radius'  \cite{Polyakov:2018zvc}\footnote{The  radius associated with the energy density $\langle r^2\rangle \sim \int d^3x \, r^2 T^{00}$ has been computed in \cite{Suganuma:2020jng}  in the present model using the numerical solution of the Yang-Mills equation.} 
\beq
\langle r^2\rangle \equiv \frac{6D(0)}{\int_0^\infty dk^2 D(k^2)}.
\eeq
Since the entire region in $k$ is involved, a proper evaluation of the  mechanical radius must include the glueball degrees of freedom.  The calculation is significantly more complex than that  of the electromagnetic form factors and the associated `charge radius' in this model \cite{Hashimoto:2008zw}  which only requires the knowledge of the asymptotic behavior $z\to \infty$ of the bulk gauge fields.   
There are other interesting directions such as the effect of finite quark masses (finite pion mass) and  generalizations to mesons and excited baryons \cite{Hayashi:2020ipd}, and perhaps even to atomic nuclei   \cite{Hashimoto:2019wmg}. We hope to address these issues in future.

\section*{Acknowledgements}
We thank Hayato Kanno for discussions. M.~F. thanks Igor Ivanov, Jiangming Yao, and Pengming Zhang for discussions. 
M.~F. and Y.~H. thank Yukawa Institute for Theoretical physics, where this work was initiated, for hospitality.  The work by T.~U. is in part supported by JSPS KAKENHI Grant Numbers 19K03831, 21K03583 and 22K03604. The work by Y.~H. is supported by the U.S. Department of Energy, Office of Science, Office of Nuclear Physics, under contract number DE- SC0012704, and also by  Laboratory Directed Research and Development (LDRD) funds from Brookhaven Science Associates. The work of S.~S is supported by 
JSPS KAKENHI (Grant-in-Aid for Scientific Research (B)) grant number JP19H01897 and MEXT KAKENHI (Grant-in-Aid for Transformative Research Areas A ``Extreme Universe") grant number 21H05187.

\end{document}